\documentclass[a4paper,11pt]{article}
\pdfoutput=1
\usepackage{jheppub}
\usepackage{booktabs}
\allowdisplaybreaks

\title{Thermodynamics of hard wall model of AdS/QCD at finite chemical potential
with Gauss--Bonnet corrections}

\author{Dharm Veer Singh,}
\author{and Shobhit Sachan\footnote{Corresponding author}}
\affiliation{Department of Physics, GLA University, Mathura, India}
\emailAdd{veerdsingh@gmail.com}
\emailAdd{shobhitsachan@gmail.com}

\abstract{We study the confinement/deconfinement transition of the hard wall model of
AdS/QCD at finite quark density in five-dimensional Einstein--Maxwell--Gauss--Bonnet
gravity, with the back-reaction of the quark matter on the geometry included. The
deconfined phase is a charged Gauss--Bonnet black hole, and the confined phase is
thermally charged AdS cut off in the infrared. The infrared value of the gauge field,
which Lee, Park and Sin fixed by requiring a consistent quark number, becomes a
function of the Gauss--Bonnet coupling. We obtain the transition line in closed form
at fixed quark chemical potential and at fixed quark number density. At fixed density
the transition passes through a region where the two phases coexist. A negative
Gauss--Bonnet coupling raises the transition temperature and extends the confined
phase to higher density. A value inside the causality bound gives a transition
temperature at zero density close to the lattice value. In the dense confined phase we compute the masses
and decay constants of the $\rho$ meson and its radial excitations. All masses increase
with density. A WKB analysis relates the
increase to a shorter effective size of the confining region, and at nuclear density
the $\rho$ mass rises by about one percent.}

\keywords{Gauge-Gravity Correspondence, Holography and Quark-Gluon Plasmas,\\
Holography and Hadronic Physics, Phase Diagram of QCD}

\begin{document}
\maketitle
\flushbottom

%==================================================================
\section{Introduction}
%==================================================================

A few microseconds after the Big Bang the universe was filled with a quark--gluon
plasma (QGP). As it cooled, the quarks and gluons combined into hadrons. Heavy-ion
collisions at the Relativistic Heavy Ion Collider (RHIC)~\cite{Back:2004je} and at the Large Hadron Collider (LHC)~\cite{Alberico:2011zy,Emerick:2011xu}
recreate this plasma. The data show that the QGP near the deconfinement temperature
behaves as a strongly coupled, nearly perfect fluid~\cite{Shuryak:2004cy,Policastro:2001yc}, so perturbative QCD does not describe it. Lattice
QCD works well at zero baryon density, where it finds a smooth crossover at
$T_{pc}=156.5\pm1.5$~MeV~\cite{HotQCD:2018pds} and determines the curvature of the
crossover line at small chemical potential~\cite{Borsanyi:2020fev}. At finite density
the fermion sign problem prevents direct simulations~\cite{Kogut:2004su},
and the phase structure at large chemical potential is still unknown.

The AdS/CFT correspondence, or gauge/gravity
duality~\cite{Maldacena:1997re,Witten:1998qj,Gubser:1998bc}, gives a different way to
study strongly coupled gauge theories. Bottom-up models of QCD built on it reproduce
many properties of hadrons~\cite{Erlich:2005qh,DaRold:2005mxj,deTeramond:2005su,Sakai:2004cn,Horigome:2006xu,Erdmenger:2007cm} and of the strongly coupled
plasma~\cite{Janik:2005zt,Herzog:2006gh,Gubser:2006bz}. The hadronic phase at finite density is less well
understood, and the phase diagrams obtained in these models differ from that expected
for QCD~\cite{Horigome:2006xu,Bergman:2007wp}. One reason is that most early studies
treated the quark matter as a probe and neglected its back-reaction on the geometry.

In the hard wall model confinement is implemented by an infrared (IR) cutoff $z_{\rm IR}$.
Herzog showed that the deconfinement transition is then dual to a Hawking--Page
transition~\cite{Hawking:1982dh,Witten:1998zw} between thermal AdS and the AdS black
hole~\cite{Herzog:2006ra}; see also~\cite{BallonBayona:2007vp,Cai:2007zw}. Kim, Lee,
Nam, Park and Sin added quark matter~\cite{Kim:2007em}. Lee, Park and Sin then included
its back-reaction, with a charged AdS black hole for the deconfined phase and a
thermally charged AdS (tcAdS) geometry for the confined phase~\cite{Lee:2009bya}. They
also computed the transition in both the grand canonical and the canonical ensemble
and the masses of vector mesons in dense matter. The soft wall version of this
construction was studied in ref.~\cite{Park:2011qq}.

In earlier work we studied the confinement\slash deconfinement transition of the soft wall
model at finite chemical potential~\cite{Sachan:2011zz}, with generalized warp
factors~\cite{Sachan:2014vxa}, and in Einstein--Maxwell--Gauss--Bonnet
gravity~\cite{Sachan:2013jpa}. There we computed
the Hawking--Page transition between the charged AdS black hole and thermally charged
AdS, and the entropy of the two phases. Holographic models of the QCD phase diagram have
since been developed further. Einstein--Maxwell--dilaton models fitted to lattice data
reproduce the equation of state and predict a critical end
point~\cite{DeWolfe:2010he,Knaute:2017opk,Cai:2022omk,Rougemont:2023gfz}. Gauss--Bonnet corrections have recently been added
to these models to obtain a temperature-dependent shear viscosity~\cite{Li:2025GB,Tong:2026GB}.

In this paper we add the Gauss--Bonnet (GB) term, the simplest higher-curvature
correction, to the back-reacted hard wall model of ref.~\cite{Lee:2009bya}. The GB term
keeps the equations of motion second order. In the dual field theory it makes the
central charges $a$ and $c$ unequal, and its coupling is bounded by causality and
positivity of energy flux~\cite{Brigante:2007nu,Brigante:2008gz,Hofman:2008ar,Buchel:2009tt,Hofman:2009ug}. The correction modifies the metric function of the
charged black hole. We compute how the Hawking temperature, the free energies and the
entropy depend on the GB coupling, and how the coupling changes the transition at fixed
quark chemical potential and at fixed quark number density. We also compute the masses
of the $\rho$ meson and its radial excitations in the dense confined phase.

The paper is organized as follows. Section~\ref{sec:GB} reviews the charged GB black
hole. Section~\ref{sec:HP} derives the Hawking temperature and the reduced on-shell
action. Sections~\ref{sec:BH} and~\ref{sec:tcAdS} compute the renormalized actions of
the charged black hole and of thermally charged AdS. Section~\ref{sec:transition}
gives the phase transition in the two ensembles, and
section~\ref{sec:mesons} gives the vector meson spectrum. Section~\ref{sec:conclusion} summarizes the results.

%==================================================================
\section{Black hole solution with Gauss--Bonnet corrections}
\label{sec:GB}
%==================================================================

We briefly review Einstein--Maxwell--Gauss--Bonnet gravity in five dimensions
with a negative cosmological constant~\cite{Cvetic:2001bk,Cai:2001dz}.
Gauss--Bonnet gravity is a special case of Lovelock
gravity~\cite{Lovelock:1971yv}, whose Lagrangian is
\begin{equation}
\mathcal{L}=\sqrt{g}\left[\alpha_0+\alpha_1R+\alpha_2\left(R^2-4R_{\mu\nu}R^{\mu\nu}
+R_{\mu\nu\rho\sigma}R^{\mu\nu\rho\sigma}\right)+\alpha_3\,\mathcal{O}(R^3)\right],
\label{eq:lovelock}
\end{equation}
where $\alpha_0$ is related to the cosmological constant $\Lambda$ and the
couplings $\alpha_n$ with $n\ge2$ parametrize corrections to Einstein gravity. In
five-dimensional Euclidean space, the most general action with curvature-squared
corrections is
\begin{equation}
S=-\int d^5x\sqrt{g}\,\frac{1}{2\kappa^2}\Big(R-2\Lambda+a\,R^2
+b\,R_{\mu\nu}R^{\mu\nu}+c\,R_{\mu\nu\rho\sigma}R^{\mu\nu\rho\sigma}\Big)
+S_{\rm matter},
\label{eq:general}
\end{equation}
where the coefficients $a$, $b$ and $c$ depend on the string compactification,
$\kappa^2=8\pi G_5$, and $\Lambda=-6$ in units where the AdS radius is $L=1$. The
choice $a=\alpha$, $b=-4\alpha$, $c=\alpha$ gives the Gauss--Bonnet combination,
which is the unique quadratic term for which the equations of motion remain second
order in derivatives. As a consequence the theory is free of ghosts when expanded
around flat space~\cite{Boulware:1985wk,Zwiebach:1985uq}. With the Maxwell field
as matter, the action is
\begin{equation}
S=-\int d^5x\sqrt{g}\left[\frac{1}{2\kappa^2}\Big(R-2\Lambda+\alpha\,
\mathcal{L}_{GB}\Big)-\frac{1}{4g^2}F^2\right],
\label{eq:action}
\end{equation}
\begin{equation}
\mathcal{L}_{GB}=R^2-4R_{\mu\nu}R^{\mu\nu}+R_{\mu\nu\rho\sigma}R^{\mu\nu\rho\sigma},
\end{equation}
where $F^2=F_{\mu\nu}F^{\mu\nu}$ and Greek indices run from 0 to 4.

Varying the metric gives
\begin{equation}
G_{\mu\nu}+\alpha H_{\mu\nu}=\frac{\kappa^2}{g^2}\,T_{\mu\nu},
\label{eq:EOM}
\end{equation}
with
\begin{align}
H_{\mu\nu}&=2\left(RR_{\mu\nu}-2R_{\mu\sigma}R^{\sigma}{}_{\nu}
-2R^{\rho\sigma}R_{\mu\rho\nu\sigma}+R_{\mu}{}^{\rho\sigma\eta}R_{\nu\rho\sigma\eta}\right)
-\frac12 g_{\mu\nu}\,\mathcal{L}_{GB},\\
G_{\mu\nu}&=R_{\mu\nu}-\frac12 Rg_{\mu\nu}+\Lambda g_{\mu\nu},\\
T_{\mu\nu}&=F_{\rho\mu}F^{\rho}{}_{\nu}-\frac14 g_{\mu\nu}F^{\rho\sigma}F_{\rho\sigma}.
\end{align}
Because the Gauss--Bonnet term is topological in four dimensions and contributes
only second-order terms to the field equations in five dimensions,
eq.~\eqref{eq:EOM} contains no derivatives of the curvature tensor. This is the
origin of the absence of ghosts noted above.

Varying the gauge field gives
\begin{equation}
\frac{1}{\sqrt g}\,\partial_\mu\left(\sqrt g\,F^{\mu\nu}\right)=0.
\end{equation}
We take only the temporal component to be non-zero and to depend on $z$ alone,
$A_t=A_t(z)$, $A_i=A_z=0$ $(i=1,2,3)$. The general solution is
\begin{equation}
A_t(z)=i\left(\mu-Qz^2\right),
\label{eq:At}
\end{equation}
where the factor $i$ arises from the Euclidean signature. The boundary is at
$z=0$, so the boundary value of $A_t$ is $i\mu$. The constants $\mu$ and $Q$ are
identified with the quark chemical potential and a quantity proportional to the
quark number density in the dual theory~\cite{Kim:2007em,Lee:2009bya}.

We take the metric ansatz
\begin{equation}
ds^2=\frac{L^2}{z^2}\left(A^2f(z)\,dt^2+\frac{dz^2}{f(z)}+\sum_{i=1}^3dx_i^2\right),
\label{eq:metric}
\end{equation}
and set $L=1$ from now on. Solving eq.~\eqref{eq:EOM} gives
\begin{equation}
f(z)=\frac{1}{4\alpha}\left(1-\sqrt{1-8\alpha\left(1-mz^4+q^2z^6\right)}\right).
\label{eq:f}
\end{equation}
The constant $A$ is fixed by requiring the boundary metric to be flat in the time
coordinate $t$:
\begin{equation}
A^2=\frac{1+\sqrt{1-8\alpha}}{2},\qquad
f_{\rm AdS}\equiv\lim_{z\to0}f(z)=\frac{1-\sqrt{1-8\alpha}}{4\alpha}=\frac{1}{A^2}.
\label{eq:A}
\end{equation}

The GB coupling is constrained by consistency of the dual field theory. The trace
anomaly of a four-dimensional CFT in a curved background is characterized by two
central charges $a$ and $c$~\cite{Duff:1977ay},
\begin{equation}
T^{\mu}{}_{\mu}=\frac{c}{16\pi^2}\,C_{\mu\nu\rho\sigma}C^{\mu\nu\rho\sigma}
-\frac{a}{16\pi^2}\left(R^2-4R_{\mu\nu}R^{\mu\nu}+R_{\mu\nu\rho\sigma}R^{\mu\nu\rho\sigma}\right),
\end{equation}
where $C_{\mu\nu\rho\sigma}$ is the Weyl tensor. The GB coupling makes $a\neq c$.
Requiring positivity of the energy flux measured by a calorimeter at null
infinity~\cite{Hofman:2008ar,Hofman:2009ug}, which is equivalent to the absence of
causality violation in the bulk plasma~\cite{Brigante:2007nu,Brigante:2008gz,Buchel:2009tt}, restricts the coupling to
\begin{equation}
-\frac{7}{72}\le\alpha\le\frac{9}{200}.
\label{eq:bound}
\end{equation}
In the conventional normalization $\lambda_{GB}=2\alpha$ this is the familiar range
$-7/36\le\lambda_{GB}\le9/100$. A finite GB coupling in a UV-complete theory also
requires an infinite tower of higher-spin states~\cite{Camanho:2014apa}, so we
treat $\alpha$ as an effective parameter. In the limit $\alpha\to0$ we recover the
Reissner--Nordstr\"om AdS (RN-AdS) metric function $f(z)=1-mz^4+q^2z^6$.

%==================================================================
\section{Dual geometry and Hawking--Page transition}
\label{sec:HP}
%==================================================================

The Hawking--Page transition is dual to the confinement\slash deconfinement
transition, and it is convenient to work in Euclidean signature. Let $z_{\rm H}$ denote
the outer horizon of eq.~\eqref{eq:f}. The horizon condition $f(z_{\rm H})=0$ gives
\begin{equation}
m=\frac{1}{z_{\rm H}^4}+q^2z_{\rm H}^2 ,
\label{eq:m}
\end{equation}
which expresses the mass parameter in terms of the charge and the horizon.

Regularity of the Euclidean geometry at the horizon fixes the period of the
Euclidean time to be $\beta=4\pi/(A|f'(z_{\rm H})|)$. The absolute value appears because
$f$ decreases towards the horizon in the $z$ coordinate. Since $f'(z_{\rm H})=-4mz_{\rm H}^3+6q^2z_{\rm H}^5$, the
Hawking temperature is
\begin{equation}
T_{\rm H}=\frac{A\,|f'(z_{\rm H})|}{4\pi}=\frac{A}{\pi z_{\rm H}}\left(1-\frac12q^2z_{\rm H}^6\right).
\label{eq:TH}
\end{equation}
Because $A_t(z)$ must be regular at the horizon, we impose the Dirichlet condition
$A_t(z_{\rm H})=0$, which gives
\begin{equation}
Q=\frac{\mu}{z_{\rm H}^2}.
\label{eq:Qmu}
\end{equation}

The relation between $q$ and $Q$ follows from the $tt$ component of
eq.~\eqref{eq:EOM}. Inverting eq.~\eqref{eq:f} gives
$f-2\alpha f^2=1-mz^4+q^2z^6$. For planar symmetry the geometric side of the $tt$
equation is a total derivative of this combination,
\begin{equation}
\left(G+\alpha H\right)^t{}_t=-\frac{3z^5}{2}\frac{d}{dz}
\left[\frac{f-2\alpha f^2}{z^4}\right]-6=-3q^2z^6 ,
\end{equation}
where the last step uses $\left(f-2\alpha f^2\right)/z^4=z^{-4}-m+q^2z^2$. The matter side is
$(\kappa^2/g^2)T^t{}_t=-2\kappa^2Q^2z^6/(g^2A^2)$. The
factor $1/A^2$ arises from $g^{tt}g^{zz}=z^4/A^2$. The GB coupling affects only
the geometric side, and we obtain
\begin{equation}
q^2=\frac23\,\frac{\kappa^2}{g^2}\,\frac{Q^2}{A^2},
\label{eq:qQ}
\end{equation}
which reduces to the result of ref.~\cite{Lee:2009bya} at $\alpha=0$; see also
ref.~\cite{Ge:2008ni}. Using eq.~\eqref{eq:Qmu},
\begin{equation}
q^2=\frac23\,\frac{\kappa^2}{g^2}\,\frac{\mu^2}{A^2z_{\rm H}^4}.
\label{eq:qmu}
\end{equation}

The holographic dictionary~\cite{Sin:2007ze,Kruczenski:2003uq} relates the
gravitational parameters to the numbers of colours $N_c$ and flavours $N_f$,
\begin{equation}
\frac{1}{2\kappa^2}=\frac{N_c^2}{8\pi^2},\qquad
\frac{1}{2g^2}=\frac{N_cN_f}{8\pi^2},\qquad
\frac{\kappa^2}{g^2}=\frac{N_f}{N_c}.
\label{eq:dictionary}
\end{equation}
Substituting eqs.~\eqref{eq:qmu} and~\eqref{eq:dictionary} into
eq.~\eqref{eq:TH} gives
\begin{equation}
\frac{1}{3A}\frac{N_f}{N_c}\,\mu^2z_{\rm H}^2+\pi T_Hz_{\rm H}-A=0 .
\label{eq:quadratic}
\end{equation}
Of its two roots only the positive one is physical,
\begin{equation}
z_{\rm H}=\frac{3A}{2\mu^2N_f/N_c}\left(\sqrt{\frac43\,\mu^2\frac{N_f}{N_c}
+\pi^2T_{\rm H}^2}-\pi T_{\rm H}\right).
\label{eq:zH}
\end{equation}
The horizon position thus depends on the GB coupling only through $A$. Here
$\mu$ is the quark chemical potential; the baryon chemical potential is
$\mu_B=3\mu$.

To evaluate the action~\eqref{eq:action} on shell, we contract
eq.~\eqref{eq:EOM} with $g^{\mu\nu}$. Using $g^{\mu\nu}H_{\mu\nu}=-\frac12\mathcal{L}_{GB}$
in five dimensions,
\begin{equation}
-\frac32R+5\Lambda-\frac{\alpha}{2}\,\mathcal{L}_{GB}=-\frac{\kappa^2}{4g^2}F_{\mu\nu}F^{\mu\nu}.
\label{eq:trace}
\end{equation}
Eliminating $\alpha\mathcal{L}_{GB}$ from eq.~\eqref{eq:action}, the $F^2$ terms
cancel and
\begin{equation}
S=-\int d^5x\sqrt g\left(\frac{-R+4\Lambda}{\kappa^2}\right).
\label{eq:Sreduced}
\end{equation}
The action therefore depends on $\alpha$ only through $\sqrt g$ and $R$. For the
metric~\eqref{eq:metric} the Ricci scalar is
\begin{equation}
R=-\left(z^2f''(z)-8z\,f'(z)+20f(z)\right),
\label{eq:Ricci}
\end{equation}
which gives $R=-20$ for the Schwarzschild--AdS black hole and $R=-20-2q^2z^6$ for the
RN-AdS black hole, in agreement with eq.~\eqref{eq:trace}. With $\sqrt g=A/z^5$ and
$\Lambda=-6$ the radial integrand is a total derivative, and
\begin{equation}
S=-\frac{A\beta V_3}{\kappa^2}\left[\frac{f'(z)}{z^3}+\frac{6-5f(z)}{z^4}\right]_{\epsilon}^{z_*},
\label{eq:master}
\end{equation}
where $V_3=\int d^3x$, $z_*=z_{\rm H}$ for the black hole and $z_*=z_{\rm IR}$ for thermal
AdS.

%==================================================================
\section{Charged black hole solution}
\label{sec:BH}
%==================================================================

The hard wall model gives a simple and instructive picture of the transition
between confinement and deconfinement. The analysis was first carried out by
Herzog~\cite{Herzog:2006ra}. He identified the transition with a Hawking--Page
transition between thermal AdS and the AdS black hole, both cut off at the IR wall
$z_{\rm IR}$.
It was extended to finite quark density in ref.~\cite{Kim:2007em} and to a
back-reacted charged geometry in ref.~\cite{Lee:2009bya}. Here we follow
ref.~\cite{Lee:2009bya} and include the Gauss--Bonnet correction.

Evaluating eq.~\eqref{eq:master} with eq.~\eqref{eq:f}, the on-shell action of
the charged GB black hole is
\begin{align}
S_{\rm AdSBH}&=-\int_0^\beta d\tau\int d^3x\int_\epsilon^{z_{\rm H}}dz\,\sqrt g
\left(\frac{-R-24}{\kappa^2}\right)\nonumber\\
&=\frac{A\beta V_3}{4\kappa^2}\left[\frac{1}{z^4}\left(
\frac{8\alpha\left(5-3mz^4+2q^2z^6\right)-5}{\alpha\sqrt{1-8\alpha\left(1-mz^4+q^2z^6\right)}}
+\frac5\alpha-24\right)\right]_{\epsilon}^{z_{\rm H}},
\label{eq:SBH}
\end{align}
where $\beta=1/T_{\rm H}$. The action diverges as the UV cutoff $\epsilon\to0$. We
renormalize it by background subtraction, subtracting thermal AdS with GB
corrections, whose metric function is constant,
\begin{equation}
f_{\rm tAdS}=\frac{1}{4\alpha}\left(1-\sqrt{1-8\alpha}\right)=f_{\rm AdS} ,
\end{equation}
with line element~\eqref{eq:metric}. Its action is
\begin{equation}
S_{\rm tAdS}=-V_3\int_0^{\beta_0}d\tau\int_\epsilon^{\infty}dz\,\sqrt g
\left(\frac{-R-24}{\kappa^2}\right)
=\frac{A\beta_0V_3}{4\kappa^2}\,\frac{24\alpha-5+5\sqrt{1-8\alpha}}{\alpha\,\epsilon^4}.
\label{eq:StAdS}
\end{equation}
The period $\beta_0$ is fixed by matching the proper length of the thermal circle
of the two geometries at the cutoff,
\begin{equation}
\beta_0=\beta\sqrt{\frac{f_{\rm AdSBH}(\epsilon)}{f_{\rm tAdS}}} .
\end{equation}
Expanding $f_{\rm AdSBH}(\epsilon)=f_{\rm AdS}-m\epsilon^4/\sqrt{1-8\alpha}+\mathcal{O}(\epsilon^6)$,
the renormalized action becomes
\begin{align}
\bar S_{\rm AdSBH}&=\lim_{\epsilon\to0}\left[S_{\rm AdSBH}-S_{\rm tAdS}\right]\nonumber\\
&=\frac{A\beta V_3}{4\kappa^2}\left[\frac{1}{z_{\rm H}^4}\left(
8\left(5-3mz_{\rm H}^4+2q^2z_{\rm H}^6\right)-24\right)+6m\right].
\label{eq:SBHren}
\end{align}
The subtraction term is $6m$ for every $\alpha$, because its coefficient
$(6-3f_{\rm AdS})/\big(2\sqrt{1-8\alpha}\,f_{\rm AdS}\big)$ equals $3/2$ identically.
Substituting eq.~\eqref{eq:m}, the bracket at $z_{\rm H}$ equals
$-8(1+q^2z_{\rm H}^6)/z_{\rm H}^4$, and
\begin{equation}
\bar S_{\rm AdSBH}=-\frac{A\beta V_3}{2\kappa^2z_{\rm H}^4}\left(1+q^2z_{\rm H}^6\right).
\label{eq:SBHfinal}
\end{equation}

The grand potential is related to the on-shell action by $\Omega=S_{\rm on\text{-}shell}/\beta$.
Per unit volume, and using eq.~\eqref{eq:qmu},
\begin{equation}
\Omega_{\rm AdSBH}=-\frac{A}{2\kappa^2z_{\rm H}^4}\left(1+q^2z_{\rm H}^6\right)
=-\frac{A}{2\kappa^2z_{\rm H}^4}-\frac{\mu^2}{3Ag^2z_{\rm H}^2}.
\label{eq:OmegaBH}
\end{equation}

Differentiating eq.~\eqref{eq:OmegaBH} at fixed $T$ and using
eq.~\eqref{eq:quadratic} to eliminate $\partial z_{\rm H}/\partial\mu$, we find
\begin{equation}
n=-\frac{\partial\Omega_{\rm AdSBH}}{\partial\mu}\bigg|_T=\frac{2Q}{Ag^2},\qquad
s=-\frac{\partial\Omega_{\rm AdSBH}}{\partial T}\bigg|_\mu=\frac{2\pi}{\kappa^2z_{\rm H}^3}
=\frac{N_c^2}{2\pi z_{\rm H}^3}.
\label{eq:first-law}
\end{equation}
The density agrees with the value obtained from the boundary term in
section~\ref{sec:bc}. The entropy equals the horizon area divided by $4G_5$, which
is the Wald entropy of a planar Gauss--Bonnet horizon. The background-subtracted
action thus gives a consistent grand canonical ensemble at finite $\mu$. At
$\mu=0$, eq.~\eqref{eq:OmegaBH} becomes $\Omega_{\rm AdSBH}=-\pi^4T^4/(2\kappa^2A^3)$.

\subsection{Entropy and equation of state of the deconfined phase}
\label{sec:entropy}

The entropy density~\eqref{eq:first-law} is the Bekenstein--Hawking entropy of the
planar horizon. The horizon area per unit boundary volume is $1/z_{\rm H}^3$, and
$1/(4G_5)=2\pi/\kappa^2$. With the dictionary~\eqref{eq:dictionary},
\begin{equation}
s=\frac{N_c^2}{2\pi z_{\rm H}^3},\qquad
\frac{s}{T^3}=\frac{\pi^2N_c^2}{2A^3}\qquad(\mu=0).
\label{eq:sT3}
\end{equation}
At zero density the entropy scales as $T^3$, as it must for a conformal plasma, and
the GB coupling enters only through $A^{-3}$. For $N_c=3$ the coefficient is
$35.2$ at $\alpha=-7/72$, $44.4$ at $\alpha=0$ and $51.0$ at $\alpha=0.04$. Negative
$\alpha$ lowers the entropy at a given temperature.

At finite $\mu$ the horizon follows from eq.~\eqref{eq:zH}. At fixed $T$ the horizon
moves towards the boundary as $\mu$ grows, so $s/T^3$ increases with $\mu/T$. It
returns to the value~\eqref{eq:sT3} for $\mu/T\to0$. The right panel of
figure~\ref{fig:dOmega} shows $s/T^3$ as a function of $T$ at $\mu=400$~MeV.

The pressure is $P=-\Omega_{\rm AdSBH}$, and the energy density is $\epsilon=\Omega+Ts+\mu n$.
Using eqs.~\eqref{eq:OmegaBH} and~\eqref{eq:first-law} we find
\begin{equation}
\epsilon=\frac{3A}{2\kappa^2z_{\rm H}^4}+\frac{\mu^2}{Ag^2z_{\rm H}^2}=3P .
\end{equation}
The trace of the stress tensor vanishes, as it should for the conformal plasma of the
black hole. The quark number density of the deconfined phase is
$n_{\rm AdSBH}=2\mu/(Ag^2z_{\rm H}^2)$. At $\mu=0$ its derivative gives the quark number
susceptibility,
\begin{equation}
\frac{\chi_{\rm AdSBH}}{T^2}=\frac{1}{T^2}\frac{\partial n_{\rm AdSBH}}{\partial\mu}\bigg|_{\mu=0}=\frac{N_cN_f}{2A^3}.
\label{eq:chiBH}
\end{equation}
This gives $\chi_{\rm AdSBH}/T^2=2.38$, $3$ and $3.44$ for $\alpha=-7/72$, $0$ and $0.04$. For
free quarks the value is $N_cN_f/3=2$.

%==================================================================
\section{Thermally charged AdS solution}
\label{sec:tcAdS}
%==================================================================

The hadronic, or confined, phase is the low-temperature phase of QCD. In the dual theory
it is described by a thermally charged AdS (tcAdS) geometry~\cite{Park:2011qq,Lee:2009bya}. The geometry ends at an IR cutoff $z_{\rm IR}$, which produces
confinement~\cite{Herzog:2006ra}, and the bulk gauge field describes the quark matter.
The tcAdS metric with Gauss--Bonnet corrections is
\begin{equation}
ds^2=\frac{L^2}{z^2}\left(A^2f(z)\,dt^2+\frac{dz^2}{f(z)}+\sum_{i=1}^3dx_i^2\right),\qquad
f(z)=\frac{1}{4\alpha}\left(1-\sqrt{1-8\alpha\left(1+q'^2z^6\right)}\right),
\label{eq:tcAdS}
\end{equation}
with $L=1$. It follows from eq.~\eqref{eq:f} with $m=0$. The solution has no horizon,
so the Euclidean time period $\beta'$ is arbitrary and tcAdS exists at every
temperature. For $\alpha\to0$ it reduces to the thermally charged AdS geometry of
ref.~\cite{Kim:2007em}, and for $q'=0$ to thermal AdS with GB corrections. For
$\alpha>0$ the metric is real only if $8\alpha(1+q'^2z_{\rm IR}^6)\le1$. The gauge field
again takes the form~\eqref{eq:At}.

\subsection{Boundary conditions and on-shell action}

For a fixed chemical potential we impose the Dirichlet condition $\mathcal A_t=i\mu$ at
the UV boundary. Following ref.~\cite{Lee:2009bya} we also impose a Dirichlet condition
at the IR cutoff,
\begin{equation}
\mathcal A_t(z_{\rm IR})=i\zeta\mu ,
\label{eq:IRbc}
\end{equation}
where $\zeta$ is a constant that we fix below. With the general solution~\eqref{eq:At} this
gives
\begin{equation}
Q=\frac{(1-\zeta)\,\mu}{z_{\rm IR}^2}.
\label{eq:Qxi}
\end{equation}
The charge parameter $q'$ in the metric is related to $Q$ by the same $tt$ equation as in
the black hole, $q'^2=\frac23\frac{\kappa^2}{g^2}\frac{Q^2}{A^2}$.

Evaluating eq.~\eqref{eq:master} between the UV cutoff $\epsilon$ and $z_{\rm IR}$ gives
\begin{equation}
S_{\rm tcAdS}=\frac{A\beta'V_3}{4\kappa^2}\left[\frac{1}{z^4}\left(
\frac{-5+8\alpha\left(5+2q'^2z^6\right)}{\alpha\sqrt{1-8\alpha\left(1+q'^2z^6\right)}}
+\frac5\alpha-24\right)\right]_{\epsilon}^{z_{\rm IR}} .
\label{eq:StcAdS}
\end{equation}
We subtract thermal AdS, eq.~\eqref{eq:StAdS}, and match the periods with
$\beta_0=\beta'\sqrt{f_{\rm tcAdS}(\epsilon)/f_{\rm tAdS}}$. Since
$f_{\rm tcAdS}(\epsilon)=f_{\rm AdS}+\mathcal O(\epsilon^6)$, the matching correction vanishes as
$\epsilon\to0$, and the grand potential per unit volume is
\begin{equation}
\Omega_{\rm tcAdS}=\frac{A}{4\kappa^2z_{\rm IR}^4}\left(
\frac{-5+8\alpha\left(5+2q'^2z_{\rm IR}^6\right)}{\alpha\sqrt{1-8\alpha\left(1+q'^2z_{\rm IR}^6\right)}}
+\frac5\alpha-24\right).
\label{eq:OmegatcAdS}
\end{equation}
For $\alpha\to0$ it reduces to $\Omega_{\rm tcAdS}=-(1+q'^2z_{\rm IR}^6)/(\kappa^2z_{\rm IR}^4)$. Expanding
in the charge we find
\begin{equation}
\Omega_{\rm tcAdS}=-\frac{6A-5A^{-1}}{\kappa^2z_{\rm IR}^4}-\frac{A\,q'^2z_{\rm IR}^2}{\sqrt{1-8\alpha}\kappa^2}
+\mathcal O(q'^4)
=-\frac{6A-5A^{-1}}{\kappa^2z_{\rm IR}^4}-\frac{2}{3\sqrt{1-8\alpha}}\,\frac{Q^2z_{\rm IR}^2}{Ag^2}
+\mathcal O(Q^4).
\label{eq:OmegatcAdS2}
\end{equation}
The first term is the vacuum energy of thermal AdS with GB corrections. The second term
is exact at $\alpha=0$.

\subsection{Quark number and the value of $\zeta$}
\label{sec:bc}

Applying a Neumann rather than a Dirichlet condition at the UV cutoff is equivalent to
a Legendre transformation from the grand potential to the free
energy~\cite{Lee:2009bya}. The required boundary term is
\begin{equation}
S_b=-\frac{1}{g^2}\int_{\partial M}d^4x\,\sqrt{g^{(4)}}\;\eta^m\mathcal A_lF_{mn}\,g^{ln},
\label{eq:Sb}
\end{equation}
where $\eta^m=(0,0,0,0,z\sqrt{f}/L)$ is the outward unit normal and the boundary
metric is
\begin{equation}
g^{(4)}={\rm diag}\left(\frac{A^2L^2f}{z^2},\frac{L^2}{z^2},\frac{L^2}{z^2},\frac{L^2}{z^2}\right),
\qquad \sqrt{g^{(4)}}=\frac{L^4A\sqrt f}{z^4}.
\end{equation} With the solution~\eqref{eq:At} the boundary
term at $z=\epsilon\to0$ is
\begin{equation}
S_b=\frac{1}{g^2}\,\frac1A\,(V_3\beta)\,(2\mu Q)=\beta\mu N,\qquad
N=\frac{2V_3Q}{Ag^2}.
\label{eq:N}
\end{equation}
The same expression holds in the black hole and in tcAdS, because the radial electric
flux $\sqrt gF^{zt}=-2iQ/A$ does not depend on $z$.

The quark number can also be computed from the bulk free energy,
\begin{equation}
N=-V_3\,\frac{\partial\Omega_{\rm tcAdS}}{\partial\mu}.
\end{equation}
With eqs.~\eqref{eq:Qxi} and~\eqref{eq:OmegatcAdS2},
\begin{equation}
N=\frac{4(1-\zeta)}{3\sqrt{1-8\alpha}}\,\frac{V_3Q}{Ag^2}.
\end{equation}
The two expressions agree only if
\begin{equation}
\zeta=1-\frac32\sqrt{1-8\alpha},
\label{eq:xi}
\end{equation}
which reduces to $\zeta=-1/2$ at $\alpha=0$, as in ref.~\cite{Lee:2009bya}. Then
$Q=\frac32\sqrt{1-8\alpha}\mu/z_{\rm IR}^2$ and
\begin{equation}
q'^2=\frac{3(1-8\alpha)}{2A^2}\,\frac{\kappa^2}{g^2}\,\frac{\mu^2}{z_{\rm IR}^4}.
\label{eq:qprime}
\end{equation}
Values of $\zeta$ are listed in table~\ref{tab:param}.

Because $\zeta$ is a constant, the condition can be imposed only at the order in $q'$ at
which it was derived. Beyond $\mathcal O(q'^2)$ the full expression~\eqref{eq:OmegatcAdS}
gives a quark number that differs from eq.~\eqref{eq:N}. For $\alpha>0$ it also has a
branch point at $8\alpha(1+q'^2z_{\rm IR}^6)=1$, where the tcAdS solution stops existing. In
the rest of the paper we therefore use the grand potential to the order at which the
quark number is consistent,
\begin{equation}
\Omega_{\rm tcAdS}(\mu)=-\frac{6A-5A^{-1}}{\kappa^2z_{\rm IR}^4}-\frac{3\sqrt{1-8\alpha}}{2}\,\frac{\mu^2}{Ag^2z_{\rm IR}^2}.
\label{eq:OmegaConf}
\end{equation}
It satisfies $-\partial\Omega_{\rm tcAdS}/\partial\mu=n$ exactly, with $n=N/V_3=2Q/(Ag^2)$.

\begin{table}[t]
\centering
\begin{tabular}{lccccc}
\toprule
$\alpha$ & $-7/72$ & $-0.05$ & $0$ & $0.02$ & $0.04$\\
\midrule
$A^2$ & 1.167 & 1.092 & 1 & 0.958 & 0.912\\
$\zeta$ & $-1.000$ & $-0.775$ & $-0.500$ & $-0.375$ & $-0.237$\\
$z_{\rm IR}^{-1}$ [MeV] & 348.2 & 336.8 & 322.4 & 315.6 & 307.9\\
\bottomrule
\end{tabular}
\caption{The constant $\zeta$ of eq.~\eqref{eq:xi} and the IR cutoff fixed by the
$\rho$-meson mass, eq.~\eqref{eq:zIR}, for several values of the GB coupling.}
\label{tab:param}
\end{table}

%==================================================================
\section{Confinement/deconfinement phase transition}
\label{sec:transition}
%==================================================================

The confinement\slash deconfinement transition of QCD is dual to the Hawking--Page
transition in the gravity sector. The deconfined phase is the charged GB black hole of
section~\ref{sec:BH}, and the confined phase is thermally charged AdS with GB
corrections and an IR cutoff, section~\ref{sec:tcAdS}. Both geometries include the
back-reaction of the quark matter. The phase with the lower free energy is the stable
one, and the transition occurs where the two free energies are equal.

We fix the IR cutoff from the mass of the $\rho$ meson (section~\ref{sec:mesons}),
\begin{equation}
\frac{1}{z_{\rm IR}}=\frac{A\,m_\rho}{\gamma_{0,1}},\qquad \gamma_{0,1}=2.405,
\label{eq:zIR}
\end{equation}
and use $N_c=3$ and $N_f=2$. The values of $z_{\rm IR}$ are listed in table~\ref{tab:param}.
The hard wall description of the deconfined phase requires the horizon to lie inside
the wall, $z_{\rm H}<z_{\rm IR}$. At $\mu=0$ this holds only for $\alpha<5/121\simeq0.041$, so we
restrict the GB coupling to $-7/72\le\alpha<5/121$. The condition is satisfied on all the transition
lines below.

\subsection{Fixed quark chemical potential}

At fixed $\mu$ the Dirichlet condition holds at the UV cutoff, and the on-shell actions
are grand potentials. The transition occurs where
\begin{equation}
\Delta\Omega(T,\mu)=\Omega_{\rm AdSBH}(T,\mu)-\Omega_{\rm tcAdS}(\mu)=0,
\end{equation}
with $\Omega_{\rm AdSBH}$ from eq.~\eqref{eq:OmegaBH}, $z_{\rm H}(T,\mu)$ from eq.~\eqref{eq:zH} and
$\Omega_{\rm tcAdS}$ from eq.~\eqref{eq:OmegaConf}. With eq.~\eqref{eq:dictionary} the
condition is a quadratic equation in $1/z_{\rm H}^2$,
\begin{equation}
\frac{A}{2z_{\rm H}^4}+\frac{N_f}{N_c}\,\frac{\mu^2}{3A\,z_{\rm H}^2}
-\frac{6A-5A^{-1}}{z_{\rm IR}^4}-\frac{N_f}{N_c}\,\frac{3\sqrt{1-8\alpha}\,\mu^2}{2A\,z_{\rm IR}^2}=0 .
\end{equation}
Its positive root gives the horizon at the transition,
\begin{equation}
\frac{1}{z_{\rm H}^2}=\frac1A\left[-\frac{N_f\mu^2}{3AN_c}
+\sqrt{\left(\frac{N_f\mu^2}{3AN_c}\right)^2+2A\left(\frac{6A-5A^{-1}}{z_{\rm IR}^4}
+\frac{3\sqrt{1-8\alpha}\,N_f\mu^2}{2AN_cz_{\rm IR}^2}\right)}\;\right],
\end{equation}
and eq.~\eqref{eq:quadratic} then gives the transition temperature in closed form,
\begin{equation}
T_c(\mu,\alpha)=\frac{1}{\pi z_{\rm H}}\left(A-\frac{N_f}{N_c}\,\frac{\mu^2z_{\rm H}^2}{3A}\right).
\label{eq:Tcmu}
\end{equation}
At $\mu=0$ it reduces to
\begin{equation}
T_c(0,\alpha)=\frac{\left[2(6A^4-5A^2)\right]^{1/4}}{\pi z_{\rm IR}}
=\frac{A\left[2(6A^4-5A^2)\right]^{1/4}}{\pi\gamma_{0,1}}\,m_\rho ,
\label{eq:Tc0}
\end{equation}
which gives Herzog's value $T_c=2^{1/4}/(\pi z_{\rm IR})=122$~MeV at
$\alpha=0$~\cite{Herzog:2006ra}.

\begin{figure}[t]
\centering
\includegraphics[width=0.7\textwidth]{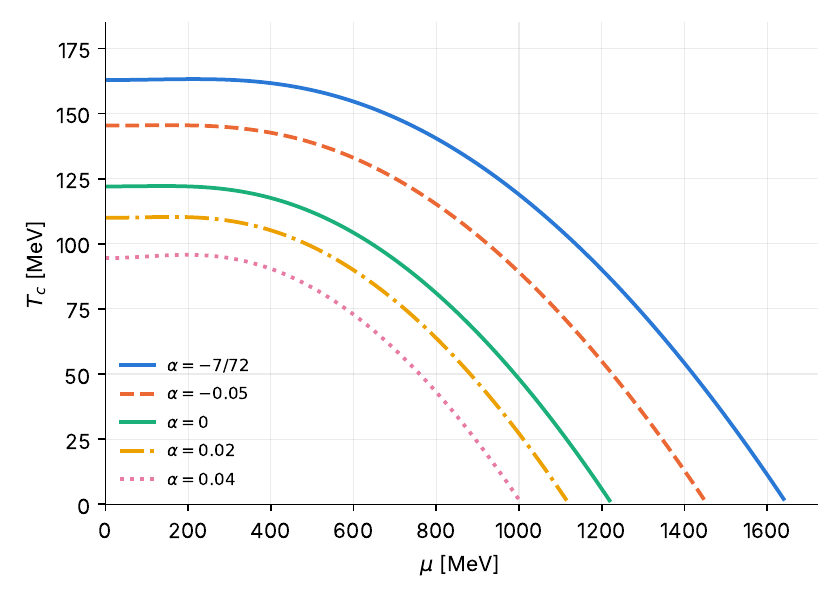}
\caption{Transition temperature at fixed quark chemical potential for several values of
the GB coupling, with $N_c=3$, $N_f=2$ and $z_{\rm IR}$ fixed by $m_\rho$.}
\label{fig:Tcmu}
\end{figure}

The transition line is shown in figure~\ref{fig:Tcmu}, and values are listed in
table~\ref{tab:Tcmu}. Negative $\alpha$ raises the transition temperature: at $\mu=0$,
$T_c$ is $163$~MeV for $\alpha=-7/72$ and $95$~MeV for $\alpha=0.04$. A coupling
$\alpha\simeq-0.08$ gives $T_c(0)\simeq157$~MeV, close to the lattice pseudo-critical
temperature~\cite{HotQCD:2018pds}.

\paragraph{Small chemical potential.}
For small $\mu$ the slope of the line follows from the Clausius--Clapeyron relation
$dT_c/d\mu=-\Delta n/\Delta s$. The entropy of tcAdS vanishes. The quark number densities
of the two phases are $n_{\rm AdSBH}=2\mu/(Ag^2z_{\rm H}^2)$ and $n_{\rm tcAdS}=3\sqrt{1-8\alpha}\mu/(Ag^2z_{\rm IR}^2)$.
Expanding eq.~\eqref{eq:Tcmu} in $\mu_B=3\mu$ gives
\begin{equation}
\frac{T_c(\mu_B)}{T_c(0)}=1-\kappa_2\,\frac{\mu_B^2}{T_c(0)^2}+\dots,\qquad
\kappa_2=\frac{N_f}{18\pi^2N_c}\left[1-\frac{3\sqrt{1-8\alpha} A}{2\sqrt{2(6A^2-5)}}\right].
\label{eq:kappa2}
\end{equation}
The bracket is negative for every allowed $\alpha$. It equals $-0.061$ at $\alpha=0$,
$-0.080$ at $\alpha=-7/72$ and $-0.21$ at $\alpha=0.04$, so $\kappa_2$ is of order
$-10^{-4}$. The transition temperature therefore first rises slightly with $\mu$, by at
most about $1.3$~MeV, because the confined phase holds more quark number than the deconfined
phase at the same $\mu$. The lattice gives $\kappa_2=0.0153$~\cite{Borsanyi:2020fev}. The
first term in the bracket is the contribution of the black hole alone. If the confined
phase carried no charge, the bracket would be $1$ and $\kappa_2=N_f/(18\pi^2N_c)\simeq0.0038$.

\paragraph{Large chemical potential.}
At larger $\mu$ the black hole gains free energy faster than tcAdS, and $T_c$ decreases.
The line reaches $T=0$ at the extremal black hole, $z_{\rm H}^2=3A^2N_c/(N_f\mu^2)$. Setting $T=0$
in eq.~\eqref{eq:Tcmu} gives
\begin{equation}
\mu_0^2z_{\rm IR}^2=\frac{A^2N_c}{2N_f}\left[9\sqrt{1-8\alpha}+\sqrt{81(1-8\alpha)+\frac{24(6A^2-5)}{A^2}}\;\right].
\end{equation}
This gives $\mu_0=1648$, $1455$, $1225$, $1121$ and $1007$~MeV for
$\alpha=-7/72$, $-0.05$, $0$, $0.02$ and $0.04$. Negative $\alpha$ extends the
confined phase to larger chemical potential.

\begin{table}[t]
\centering
\begin{tabular}{lcccccccc}
\toprule
$\alpha$ & $\mu=0$ & $200$ & $400$ & $600$ & $800$ & $1000$ & $\mu_0$\\
\midrule
$-7/72$ & 162.9 & 163.3 & 161.7 & 154.6 & 140.5 & 118.9 & 1648\\
$-0.05$ & 145.4 & 145.5 & 142.7 & 133.0 & 115.2 & \phantom{1}89.1 & 1455\\
$0$ & 122.0 & 122.0 & 117.6 & 104.2 & \phantom{1}81.0 & \phantom{1}48.1 & 1225\\
$0.02$ & 110.0 & 110.2 & 105.2 & \phantom{1}89.9 & \phantom{1}63.8 & \phantom{1}27.1 & 1121\\
$0.04$ & \phantom{1}94.5 & \phantom{1}95.8 & \phantom{1}90.4 & \phantom{1}72.8 & \phantom{1}43.0 & \phantom{11}1.7 & 1007\\
\bottomrule
\end{tabular}
\caption{Transition temperature in MeV at fixed quark chemical potential $\mu$ (in MeV),
and the chemical potential $\mu_0$ at which $T_c$ vanishes.}
\label{tab:Tcmu}
\end{table}

\begin{figure}[t]
\centering
\includegraphics[width=\textwidth]{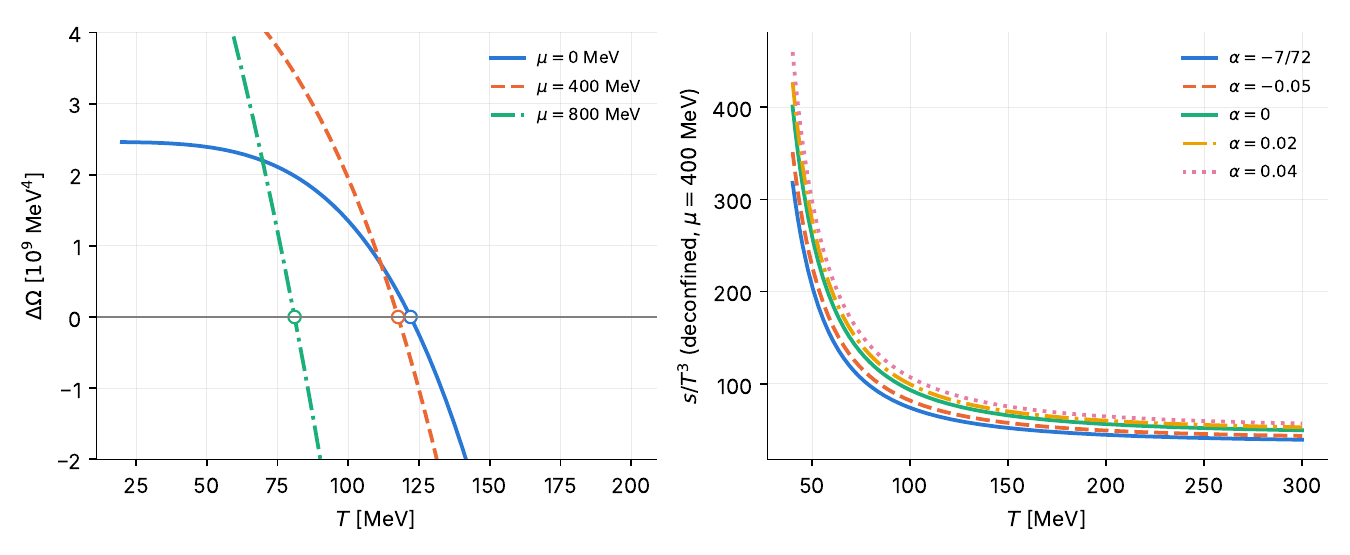}
\caption{Left: grand potential difference $\Delta\Omega=\Omega_{\rm AdSBH}-\Omega_{\rm tcAdS}$ at $\alpha=0$;
the open circles mark $T_c$. Right: entropy density of the deconfined phase divided by
$T^3$ at $\mu=400$~MeV for several values of the GB coupling.}
\label{fig:dOmega}
\end{figure}

Figure~\ref{fig:dOmega} (left) shows $\Delta\Omega(T)$ at $\alpha=0$ for three values of $\mu$.
At low temperature $\Delta\Omega>0$ and thermally charged AdS is the stable phase.
$\Delta\Omega$ decreases monotonically with $T$, since
$\partial\Delta\Omega/\partial T=-s_{\rm AdSBH}<0$, and it crosses zero once, at $T_c$. Above $T_c$
the black hole has the lower grand potential and the system is deconfined. At larger
$\mu$ the curve becomes steeper and crosses zero at a lower temperature, $T_c=122.0$,
$117.6$ and $81.0$~MeV for $\mu=0$, $400$ and $800$~MeV.

\begin{figure}[t]
\centering
\includegraphics[width=\textwidth]{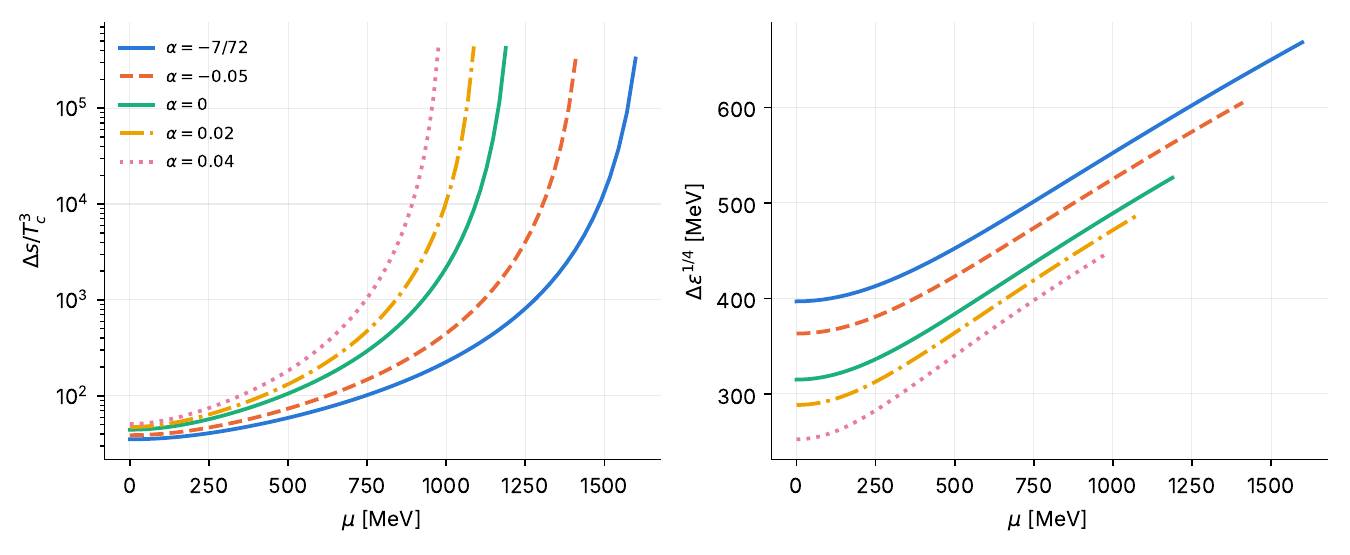}
\caption{Left: entropy jump $\Delta s/T_c^3$ along the fixed-$\mu$ transition line.
Right: latent heat $\Delta\epsilon=T_c\Delta s+\mu\Delta n$.}
\label{fig:jumps}
\end{figure}

\paragraph{Entropy, density and latent heat at the transition.}
The confined phase has $s=0$ at this order and quark number density
\begin{equation}
n_{\rm tcAdS}=-\frac{\partial\Omega_{\rm tcAdS}}{\partial\mu}=\frac{3\sqrt{1-8\alpha}\,\mu}{Ag^2z_{\rm IR}^2},
\end{equation}
which is linear in $\mu$ and independent of $T$. The deconfined phase has
$s_{\rm AdSBH}=N_c^2/(2\pi z_{\rm H}^3)$ and $n_{\rm AdSBH}=2\mu/(Ag^2z_{\rm H}^2)$. At $T_c$ both quantities jump.
The entropy jump $\Delta s=s_{\rm AdSBH}(T_c)$ is shown in figure~\ref{fig:jumps} (left). At $\mu=0$,
$\Delta s/T_c^3=\pi^2N_c^2/(2A^3)$. Along the line $\Delta s/T_c^3$ grows, slowly at first and
then steeply, because $z_{\rm H}$ at the transition stays finite as $T_c\to0$. At $\alpha=0$
it is $56$ at $\mu=242$~MeV, $100$ at $483$~MeV and $258$ at $725$~MeV.

\begin{figure}[t]
\centering
\includegraphics[width=0.66\textwidth]{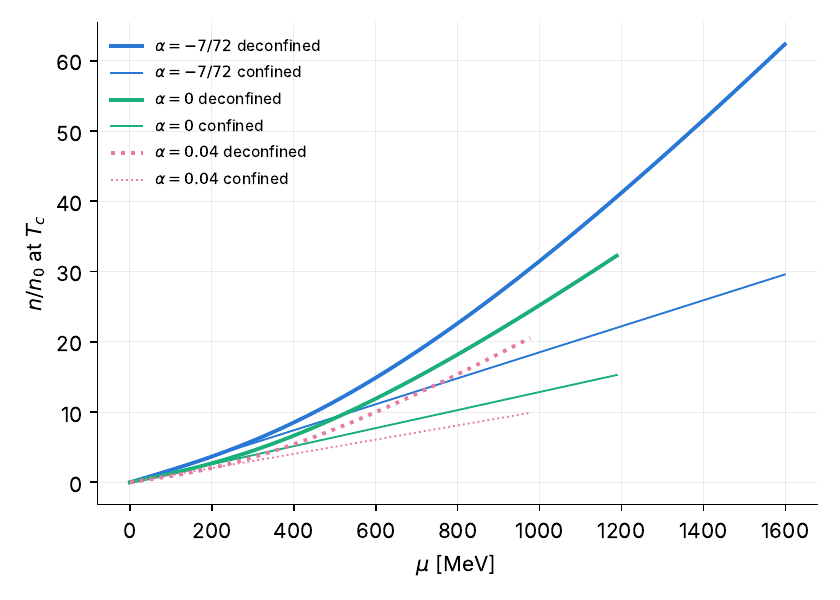}
\caption{Quark number density of the deconfined (thick lines) and confined (thin
lines) phases at the transition temperature, as functions of $\mu$, in units of the
quark density of nuclear matter $n_0=0.48~{\rm fm}^{-3}$.}
\label{fig:nmu}
\end{figure}

The quark number densities of the two phases along the transition line are shown in
figure~\ref{fig:nmu}. At small $\mu$ the confined phase holds more quark number,
$n_{\rm tcAdS}>n_{\rm AdSBH}$. By the Clausius--Clapeyron relation $dT_c/d\mu=-\Delta n/\Delta s$ this
is why $T_c$ rises at first. The two densities become equal at the maximum of $T_c$, at
$\mu\simeq140$~MeV for $\alpha=0$ and $213$~MeV for $\alpha=-7/72$. Beyond that the black hole holds more charge
and $T_c$ decreases. We checked the Clausius--Clapeyron relation along the numerical
line: at $\alpha=0$ and $\mu=604$~MeV, $dT_c/d\mu=-0.0925$ and $-\Delta n/\Delta s=-0.0925$.

Since the pressures are equal at the transition, the jump of the energy density is
\begin{equation}
\Delta\epsilon=T_c\,\Delta s+\mu\,\Delta n ,
\end{equation}
which is the latent heat per unit volume. At $\mu=0$ it reduces to
$\Delta\epsilon=4(6A-5A^{-1})/(\kappa^2z_{\rm IR}^4)$, which gives $\Delta\epsilon^{1/4}=397$, $315$ and
$253$~MeV for $\alpha=-7/72$, $0$ and $0.04$. Figure~\ref{fig:jumps} (right) shows that
$\Delta\epsilon^{1/4}$ increases along the transition line, to about $480$~MeV at $\mu\simeq970$~MeV for $\alpha=0$.

\subsection{Fixed quark number density}

To fix the quark number density $n=N/V_3$ instead of the chemical potential, we fix the
radial electric flux at the boundary. This is a Neumann condition, and the boundary
term~\eqref{eq:Sb} must be added to the action. The result is the Helmholtz free energy,
the Legendre transform $F(T,n)=\Omega(T,\mu)+\mu n$. In both phases the charge is fixed by
the density, $Q=Ag^2n/2$, so the black hole and tcAdS carry the same charge parameter,
$q'=q$.

\paragraph{Deconfined phase.}
In the black hole $\mu=Qz_{\rm H}^2$, so
\begin{equation}
\mu n=\frac{2Q^2z_{\rm H}^2}{Ag^2}=\frac{3A}{\kappa^2}\,q^2z_{\rm H}^2 ,
\end{equation}
and
\begin{equation}
F_{\rm AdSBH}=-\frac{A}{2\kappa^2z_{\rm H}^4}\left(1+q^2z_{\rm H}^6\right)+\frac{3A}{\kappa^2}\,q^2z_{\rm H}^2
=\frac{A}{2\kappa^2}\left(-\frac{1}{z_{\rm H}^4}+5q^2z_{\rm H}^2\right)
=-\frac{A}{2\kappa^2z_{\rm H}^4}+\frac{5Ag^2}{12}\,n^2z_{\rm H}^2 .
\label{eq:FBH}
\end{equation}
At fixed $n$ the horizon follows from eq.~\eqref{eq:TH},
$\pi Tz_{\rm H}=A\left(1-\frac12q^2z_{\rm H}^6\right)$, which has a unique solution because $T$
decreases monotonically with $z_{\rm H}$. One checks that $\partial F_{\rm AdSBH}/\partial n|_T=\mu$.

\paragraph{Confined phase.}
In tcAdS, eq.~\eqref{eq:Qxi} with eq.~\eqref{eq:xi} gives $\mu=2Qz_{\rm IR}^2/(3\sqrt{1-8\alpha})$, which
reduces to $\mu=\frac23Qz_{\rm IR}^2$ at $\alpha=0$. Then
\begin{equation}
\mu n=\frac{4Q^2z_{\rm IR}^2}{3\sqrt{1-8\alpha} Ag^2}=\frac{2A}{\sqrt{1-8\alpha}\kappa^2}\,q'^2z_{\rm IR}^2 ,
\end{equation}
and the free energy, to the same order as eq.~\eqref{eq:OmegaConf}, is
\begin{equation}
F_{\rm tcAdS}=-\frac{6A-5A^{-1}}{\kappa^2z_{\rm IR}^4}+\frac{A}{\sqrt{1-8\alpha}\kappa^2}\,q'^2z_{\rm IR}^2
=-\frac{6A-5A^{-1}}{\kappa^2z_{\rm IR}^4}+\frac{Ag^2}{6\sqrt{1-8\alpha}}\,n^2z_{\rm IR}^2 .
\label{eq:FtcAdS}
\end{equation}
It satisfies $\partial F_{\rm tcAdS}/\partial n=\mu$.

\paragraph{Difference of the on-shell actions.}
Following section~3.2 of ref.~\cite{Lee:2009bya}, we compute the Neumann actions of the
two geometries with the UV cutoff kept. The black hole gives
\begin{equation}
S^N_{\rm AdSBH}=\frac{A\beta V_3}{\kappa^2}\left[\frac{6-5f_{\rm AdS}}{\epsilon^4}
+\frac{1}{\sqrt{1-8\alpha}}\left(\frac{1}{z_{\rm H}^4}+q^2z_{\rm H}^2\right)-\frac{2}{z_{\rm H}^4}+q^2z_{\rm H}^2\right],
\end{equation}
and thermally charged AdS, to order $q'^2$,
\begin{equation}
S^N_{\rm tcAdS}=\frac{A\beta'V_3}{\kappa^2}\left[\frac{6-5f_{\rm AdS}}{\epsilon^4}
-\frac{6-5f_{\rm AdS}}{z_{\rm IR}^4}+\frac{q'^2z_{\rm IR}^2}{\sqrt{1-8\alpha}}\right].
\end{equation}
At $\alpha=0$ these are eq.~(3.7) of ref.~\cite{Lee:2009bya}. We match the circumferences of
the Euclidean time circles at $z=\epsilon$,
$\beta'=\beta\big(1-m\epsilon^4/(2\sqrt{1-8\alpha}\,f_{\rm AdS})\big)$ with $\beta=1/T_{\rm H}$. The $\epsilon^{-4}$ terms then
cancel, and the $m$-dependent terms combine with the coefficient
$1/\sqrt{1-8\alpha}+(6-5f_{\rm AdS})/(2\sqrt{1-8\alpha} f_{\rm AdS})=3/2$. With $q'=q$,
\begin{equation}
\Delta S^N=\frac{A\beta V_3}{\kappa^2}\left[\frac{6-5A^{-2}}{z_{\rm IR}^4}-\frac{1}{2z_{\rm H}^4}
+\frac{\kappa^2Q^2}{3g^2A^2}\left(5z_{\rm H}^2-\frac{2z_{\rm IR}^2}{\sqrt{1-8\alpha}}\right)\right],
\label{eq:dSN}
\end{equation}
which reduces to eq.~(3.8) of ref.~\cite{Lee:2009bya} at $\alpha=0$ and agrees with
$\Delta F=F_{\rm AdSBH}-F_{\rm tcAdS}$ from eqs.~\eqref{eq:FBH} and~\eqref{eq:FtcAdS}. For small $z_{\rm H}$
the term $-1/(2z_{\rm H}^4)$ dominates and the black hole is stable.

\paragraph{Transition line in closed form.}
The transition occurs at $\Delta S^N=0$. We denote by $z_{\rm H}$ the horizon of the
AdS black hole at the transition. Solving $\Delta S^N=0$ for the quark number density and
using the temperature relation $\pi T_{\rm H}z_{\rm H}=A\big(1-\frac12q^2z_{\rm H}^6\big)$ with
eq.~\eqref{eq:qQ}, we find
\begin{equation}
Q_c^2=\frac{3N_c}{2N_f}\,
\frac{A^2z_{\rm IR}^4-2(6A^2-5)\,z_{\rm H}^4}
{z_{\rm IR}^4\,z_{\rm H}^4\left(5z_{\rm H}^2-2z_{\rm IR}^2/\sqrt{1-8\alpha}\right)},
\label{eq:Tcn}
\end{equation}
\begin{equation}
T_c=\frac{A}{\pi z_{\rm H}}\left[1-\frac{z_{\rm H}^2\left(A^2z_{\rm IR}^4-2(6A^2-5)\,z_{\rm H}^4\right)}
{2A^2z_{\rm IR}^4\left(5z_{\rm H}^2-2z_{\rm IR}^2/\sqrt{1-8\alpha}\right)}\right].
\label{eq:Tcn2}
\end{equation}
As $z_{\rm H}$ varies, eqs.~\eqref{eq:Tcn} and~\eqref{eq:Tcn2} trace the transition line,
with the density given by $n=2Q_c/(Ag^2)$. At $\alpha=0$ ($A=1$) they reduce to eq.~(3.9) of
ref.~\cite{Lee:2009bya}. Positivity of $Q_c^2$ restricts the horizon to
\begin{equation}
\frac{2z_{\rm IR}^2}{5\sqrt{1-8\alpha}}<z_{\rm H}^2,\qquad
z_{\rm H}^4\le\frac{A^2z_{\rm IR}^4}{2(6A^2-5)},
\end{equation}
that is $0.6325\,z_{\rm IR}<z_{\rm H}\le0.8409\,z_{\rm IR}$ at $\alpha=0$. The upper end is $n=0$, where
eq.~\eqref{eq:Tcn2} reproduces $T_c(0)$ of eq.~\eqref{eq:Tc0}. The line ends at $T_c=0$,
where $z_{\rm H}$ is the positive root of
\begin{equation}
2(6A^2-5)\,z_{\rm H}^6+9A^2z_{\rm IR}^4z_{\rm H}^2-\frac{4A^2z_{\rm IR}^6}{\sqrt{1-8\alpha}}=0 .
\end{equation}
At $\alpha=0$ the root is $z_{\rm H}=0.6536\,z_{\rm IR}$, so the part of the line with
$T_c\ge0$ is $0.6536\,z_{\rm IR}\le z_{\rm H}\le0.8409\,z_{\rm IR}$, narrower than the range allowed
by $Q_c^2>0$. At $\alpha=0$ and $N_f/N_c=2/3$ we find $Q_c=0.207$~GeV$^3$ at $T_c=45$~MeV, in agreement
with figure~2 of ref.~\cite{Lee:2009bya}. The closed form agrees with a direct numerical
solution of $F_{\rm AdSBH}(T,n)=F_{\rm tcAdS}(n)$ to all digits shown.

We express the density in units of the quark number density of nuclear matter,
$n_0=3\times0.16~{\rm fm}^{-3}=0.48~{\rm fm}^{-3}$. The results are shown in
figure~\ref{fig:Tcn} and table~\ref{tab:Tcn}. $T_c$ stays almost constant up to
$n\simeq4n_0$, then decreases and vanishes at $n=13.5\,n_0$ for $\alpha=0.04$, $21.0\,n_0$
for $\alpha=0$ and $40.6\,n_0$ for $\alpha=-7/72$.

\paragraph{Slope and maximum.}
Along the line $\Delta F(T_c(n),n)=0$, so $-\Delta s\,dT_c+\Delta\mu\,dn=0$ and
\begin{equation}
\frac{dT_c}{dn}=\frac{\Delta\mu}{\Delta s},\qquad
\Delta\mu=\mu_{\rm AdSBH}-\mu_{\rm tcAdS}=Q\left(z_{\rm H}^2-\frac{2z_{\rm IR}^2}{3\sqrt{1-8\alpha}}\right),\qquad
\Delta s=\frac{N_c^2}{2\pi z_{\rm H}^3}.
\end{equation}
$T_c$ therefore has its maximum exactly at $z_{\rm H}^2=2z_{\rm IR}^2/(3\sqrt{1-8\alpha})$, where
\begin{equation}
T_{c,\max}=\frac{A}{\pi z_{\rm IR}}\sqrt{\frac{3\sqrt{1-8\alpha}}{2}}
\left(\frac34+\frac{2(6A^2-5)}{9A^2(1-8\alpha)}\right).
\end{equation}
At $\alpha=0$ this is $122.19$~MeV at $n=1.79\,n_0$, only $0.16$~MeV above $T_c(0)$. Near
$n=0$ the same relation gives
\begin{equation}
\frac{T_c(n)}{T_c(0)}=1+\frac{2\pi^4}{N_c^3N_f}\,z_{\rm H}^4
\left(z_{\rm H}^2-\frac{2z_{\rm IR}^2}{3\sqrt{1-8\alpha}}\right)n^2,\qquad
z_{\rm H}^4=\frac{A^2z_{\rm IR}^4}{2(6A^2-5)},
\end{equation}
which is $T_c(n)/T_c(0)=1+c\,(n/n_0)^2$ with $c=3.2\times10^{-4}$, $8.8\times10^{-4}$ and
$9.6\times10^{-3}$ for $\alpha=-7/72$, $0$ and $0.04$.

\begin{figure}[t]
\centering
\includegraphics[width=\textwidth]{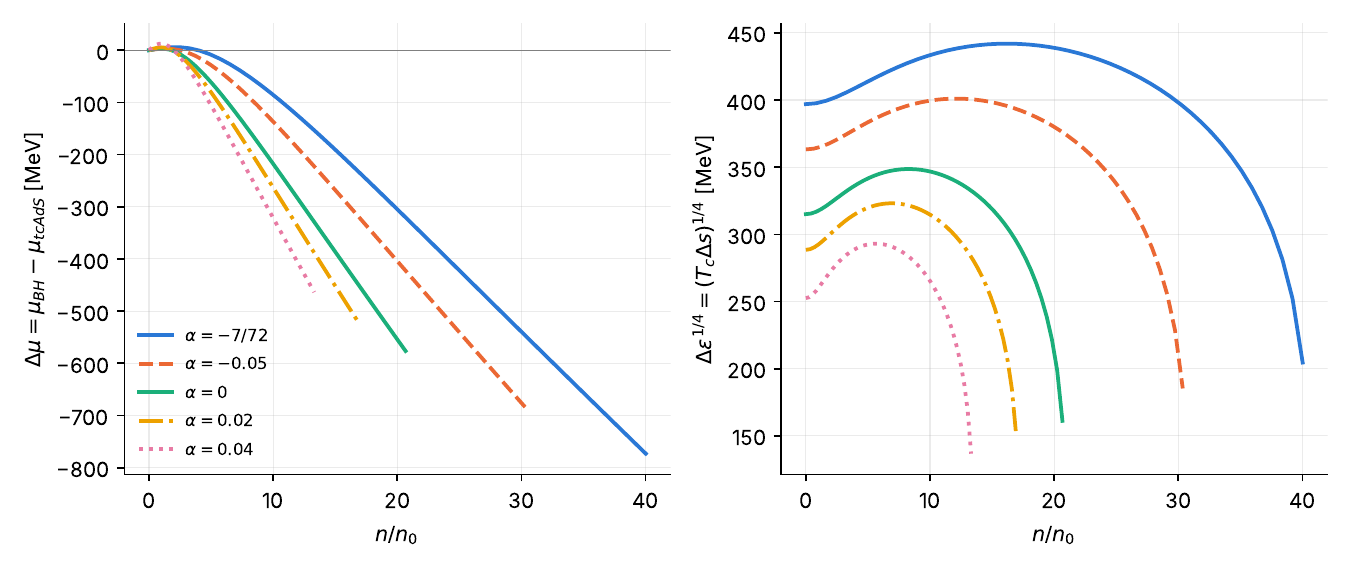}
\caption{Jump of the chemical potential (left) and latent heat (right) along the
fixed-density transition line.}
\label{fig:jumpsn}
\end{figure}

\begin{figure}[t]
\centering
\includegraphics[width=0.66\textwidth]{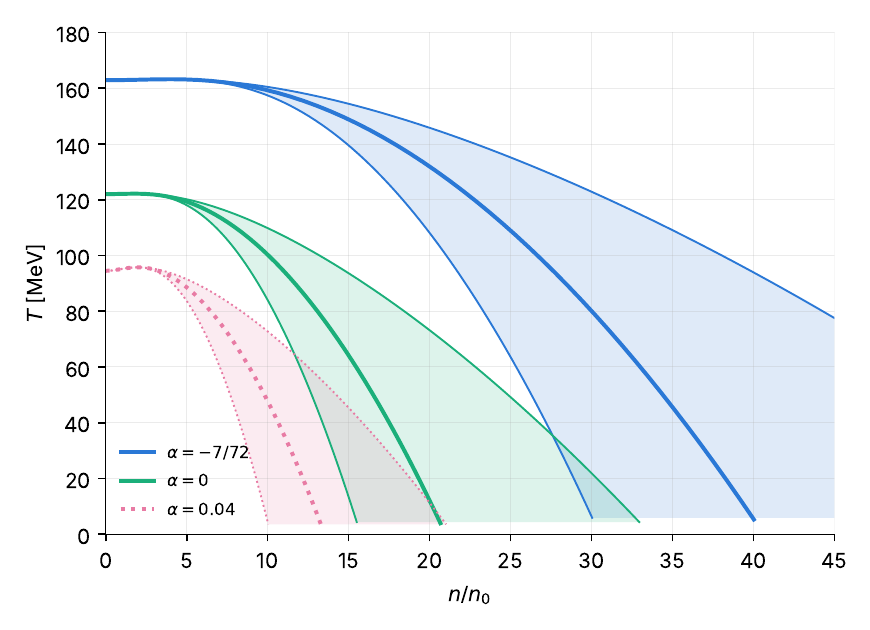}
\caption{Phase diagram at fixed quark number density. Shaded bands: coexistence region
bounded by the densities of the confined and deconfined phases on the fixed-$\mu$
transition line (thin lines). Thick lines: the equal-free-energy line,
eq.~\eqref{eq:Tcn}.}
\label{fig:coex}
\end{figure}

\begin{figure}[t]
\centering
\includegraphics[width=0.7\textwidth]{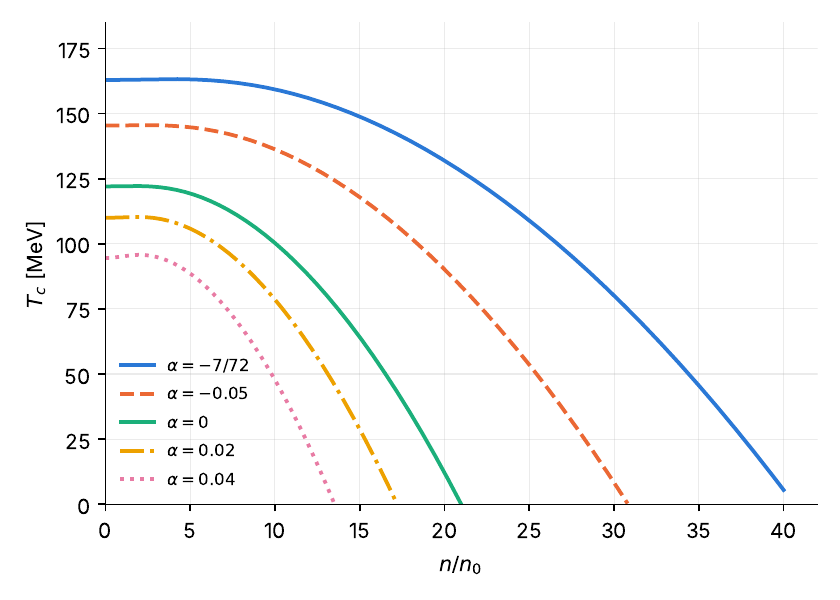}
\caption{Transition temperature at fixed quark number density, in units of the quark
density of nuclear matter $n_0=0.48~{\rm fm}^{-3}$.}
\label{fig:Tcn}
\end{figure}

\paragraph{Jumps and coexistence.}
At the transition the free energy and the density are continuous, while the entropy,
the chemical potential, the energy density and the pressure jump:
$\Delta s=N_c^2/(2\pi z_{\rm H}^3)$, $\Delta\epsilon=T_c\Delta s$ and $\Delta P=n\,\Delta\mu$. At $\alpha=0$
the latent heat $\Delta\epsilon^{1/4}$ is $318$~MeV at $n_0$ and $349$~MeV at $8n_0$, while
$\Delta\mu$ goes from $+3$~MeV at $n_0$ to $-152$~MeV at $8n_0$ (figure~\ref{fig:jumpsn}). Because
$\Delta P\neq0$, the line $F_{\rm AdSBH}=F_{\rm tcAdS}$ is not where the two phases coexist. A
first-order transition at fixed density proceeds through a mixed phase. Coexistence
requires equal $T$, $\mu$ and $P$, which is the fixed-$\mu$ condition of the previous
subsection. On that line the two phases have the densities $n_{\rm AdSBH}=2\mu/(Ag^2z_{\rm H}^2)$ and
$n_{\rm tcAdS}=3\sqrt{1-8\alpha}\mu/(Ag^2z_{\rm IR}^2)$, and every density between them is a mixture. This gives
the coexistence bands of figure~\ref{fig:coex}, which contain the equal-free-energy line.
At $\alpha=0$ and $\mu=400$~MeV, for example, the band spans $5.10\,n_0<n<6.57\,n_0$ at
$T=117.7$~MeV.

\begin{table}[t]
\centering
\small\setlength{\tabcolsep}{4pt}
\begin{tabular}{lccccccc}
\toprule
$\alpha$ & $n=0$ & $2n_0$ & $4n_0$ & $8n_0$ & $12n_0$ & $16n_0$ & $20n_0$\\
\midrule
$-7/72$ & 162.9 & 163.1 (108) & 163.3 (216) & 161.6 (432) & 155.9 (649) & 146.0 (865) & 131.9 (1081)\\
$-0.05$ & 145.4 & 145.5 (126) & 145.3 (252) & 140.9 (504) & 130.1 (756) & 113.1 (1007) & \phantom{1}90.2 (1259)\\
$0$ & 122.0 & 122.2 (156) & 121.0 (311) & 110.1 (623) & \phantom{1}87.8 (934) & \phantom{1}55.0 (1245) & \phantom{1}12.1 (1557)\\
$0.02$ & 110.0 & 110.3 (174) & 108.3 (347) & \phantom{1}92.3 (694) & \phantom{1}61.3 (1041) & \phantom{1}16.2 (1388) & --\\
$0.04$ & \phantom{1}94.5 & \phantom{1}95.8 (198) & \phantom{1}92.6 (395) & \phantom{1}68.1 (791) & \phantom{1}22.4 (1186) & -- & --\\
\bottomrule
\end{tabular}
\caption{Transition temperature in MeV at fixed quark number density. The number in
parentheses is the chemical potential of the confined phase at the transition, in MeV.
A dash means that $T_c$ has already reached zero.}
\label{tab:Tcn}
\end{table}

The tcAdS geometry has no horizon, so its entropy vanishes at the order considered. The
black hole has $s=2\pi/(\kappa^2z_{\rm H}^3)$, eq.~\eqref{eq:first-law}, in both ensembles. At
fixed $n$ the horizon is fixed by $\pi Tz_{\rm H}=A(1-\frac12q^2z_{\rm H}^6)$ with $q$ set by $n$, and
$T$ decreases monotonically with $z_{\rm H}$. The entropy thus increases with $T$, and the
heat capacity $C_n=T(\partial s/\partial T)_n$ is positive. The same holds at fixed $\mu$,
so the deconfined phase is thermally stable in both ensembles. The entropy jumps from
zero to $s_{\rm AdSBH}(T_c)$ at the transition, so the transition is of first order for
every $\alpha$ in the allowed range, at fixed $\mu$ and at fixed $n$.

%==================================================================
\section{Mass of the excited vector mesons}
\label{sec:mesons}
%==================================================================

Vector mesons are normalizable modes of a flavour gauge field $V_M$ in the confined
geometry, with action $-\frac{1}{4g_5^2}\int d^5x\sqrt{-g}\,F_{MN}F^{MN}$. We treat this
field as a probe of the tcAdS background~\eqref{eq:tcAdS}, as in ref.~\cite{Lee:2009bya}.
The density enters only through the metric. In Lorentzian signature, take a transverse
mode at rest, $V_i=\epsilon_i\,e^{-i\omega t}\psi(z)$, in the gauge $V_z=0$. With
$\sqrt{-g}=A/z^5$, $g^{zz}=z^2f$, $g^{ii}=z^2$ and $g^{tt}=-z^2/(A^2f)$, the equation of
motion is
\begin{equation}
\partial_z\!\left(\frac{f(z)}{z}\,\partial_z\psi\right)+\frac{m^2}{A^2f(z)\,z}\,\psi=0,
\label{eq:vector}
\end{equation}
with $m=\omega$ the meson mass. We impose $\psi(0)=0$, so that the mode is normalizable,
and the Neumann condition $\partial_z\psi(z_{\rm IR})=0$ of ref.~\cite{Erlich:2005qh}.

\subsection{Spectrum at zero and finite density}

\paragraph{Zero density.}
For $q'=0$ the metric function is constant, $f=f_{\rm AdS}=1/A^2$, and eq.~\eqref{eq:vector}
becomes
\begin{equation}
\psi''(z)-\frac{1}{z}\,\psi'(z)+A^2m^2\,\psi(z)=0,\qquad
\psi(0)=0,\qquad \psi'(z_{\rm IR})=0 .
\label{eq:vector0}
\end{equation}
With $\psi=z\,\chi(x)$ and $x=Amz$ it turns into Bessel's equation of order one,
$x^2\chi''+x\chi'+(x^2-1)\chi=0$. The solution regular at the boundary is
\begin{equation}
\psi(z)=z\,J_1(Amz),\qquad \psi'(z)=Amz\,J_0(Amz),
\label{eq:psi0}
\end{equation}
where we used $\frac{d}{dx}\big(xJ_1(x)\big)=xJ_0(x)$. The Neumann condition at the wall then gives
$J_0(Am_nz_{\rm IR})=0$:
\begin{equation}
m_n=\frac{\gamma_{0,n}}{A\,z_{\rm IR}},\qquad
\gamma_{0,n}=2.405,\ 5.520,\ 8.654,\ 11.792 .
\label{eq:mn0}
\end{equation}
The GB coupling enters through the factor $1/A$, the inverse of the effective AdS
radius. Fixing the lowest mode to $m_\rho=775.3$~MeV gives eq.~\eqref{eq:zIR}. The excited
states are then independent of $\alpha$,
\begin{equation}
m_2=1780~{\rm MeV},\qquad m_3=2790~{\rm MeV},\qquad m_4=3801~{\rm MeV}.
\end{equation}
The first two can be compared with the $\rho(1450)$ at $1465\pm25$~MeV and the
$\rho(1700)$ at $1720\pm20$~MeV~\cite{ParticleDataGroup:2024cfk}. As usual in the hard
wall model the excitations are too heavy. The masses grow linearly with $n$ rather than
as $m_n^2\propto n$.

\paragraph{Finite density.}
For $q'\ne0$ the metric function~\eqref{eq:tcAdS} grows towards the wall, with
\begin{equation}
f'(z)=\frac{6q'^2z^5}{\sqrt{1-8\alpha\left(1+q'^2z^6\right)}}>0 .
\end{equation}
Expanding the derivative in eq.~\eqref{eq:vector}, the equation to be solved is
\begin{equation}
\psi''(z)+\left(\frac{f'(z)}{f(z)}-\frac1z\right)\psi'(z)+\frac{m^2}{A^2f(z)^2}\,\psi(z)=0,\qquad
\psi(0)=0,\qquad \psi'(z_{\rm IR})=0 .
\label{eq:vectorn}
\end{equation}
Near the boundary $f=f_{\rm AdS}+\mathcal O(z^6)$, so eq.~\eqref{eq:vectorn} agrees with
eq.~\eqref{eq:vector0} up to order $z^6$ and the normalizable solution starts as
$\psi=z^2\big[1-A^2m^2z^2/8+\mathcal O(z^4)\big]$. At $\alpha=0$, where $f=1+q'^2z^6$,
eq.~\eqref{eq:vectorn} reads
\begin{equation}
\psi''-\frac1z\,\frac{1-5q'^2z^6}{1+q'^2z^6}\,\psi'+\frac{m^2}{\left(1+q'^2z^6\right)^2}\,\psi=0 .
\label{eq:vectorn0}
\end{equation}
The density appears both in the first-derivative term, through $f'/f$, and in the mass
term, through $1/f^2$. The latter comes from $g^{tt}=-z^2/(A^2f)$. Eq.~\eqref{eq:vectorn} has
no closed-form solution and we solve it numerically. With $y=z/z_{\rm IR}$, $M=mz_{\rm IR}$ and
$P=f\psi'/y$ it becomes the first-order system
\begin{equation}
\frac{d\psi}{dy}=\frac{y}{f}\,P,\qquad \frac{dP}{dy}=-\frac{M^2}{A^2f\,y}\,\psi ,
\label{eq:system}
\end{equation}
which we integrate from $y_0=10^{-5}$ with $\psi=y_0^2$, $P=2f(y_0)$, and the masses are the
values of $M$ for which $P(1)=0$. The density enters through the dimensionless combination
\begin{equation}
\eta\equiv q'^2z_{\rm IR}^6=\frac{2}{3}\frac{\kappa^2}{g^2}\frac{Q^2z_{\rm IR}^6}{A^2}
=\frac{8\pi^4}{3N_c^3N_f}\,n^2z_{\rm IR}^6 .
\end{equation}
For small $\eta$ first-order perturbation theory applies. Eq.~\eqref{eq:vector} is a
Sturm--Liouville problem. With $y=z/z_{\rm IR}$, the eigenvalue $m^2z_{\rm IR}^2$ is the stationary
value of
\begin{equation}
A^2\,\frac{\int_0^1dy\,f\psi'^2/y}{\int_0^1dy\,\psi^2/(fy)} .
\end{equation}
At $\alpha=0$, $\delta f=\eta y^6$ and
\begin{equation}
m_n^2=m_n^2(0)\left(1+c_n\,\eta+\mathcal O(\eta^2)\right),\qquad
c_n=\frac{\int_0^1dy\,y^5\left(\psi_n'^2+\gamma_{0,n}^2\psi_n^2\right)}{\gamma_{0,n}^2\int_0^1dy\,\psi_n^2/y},
\qquad \psi_n=y\,J_1(\gamma_{0,n}y).
\label{eq:cn}
\end{equation}
The integrals give $c_n=0.321$, $0.300$, $0.292$ and $0.289$ for $n=1,\dots,4$. The full
numerical values of $c_n(\alpha)$ are listed in table~\ref{tab:cn}. Each $c_n$ is positive, so
the density raises all the masses. The coefficient decreases slowly with $n$ towards a
limiting value, and it increases with $\alpha$.

\begin{figure}[t]
\centering
\includegraphics[width=0.82\textwidth]{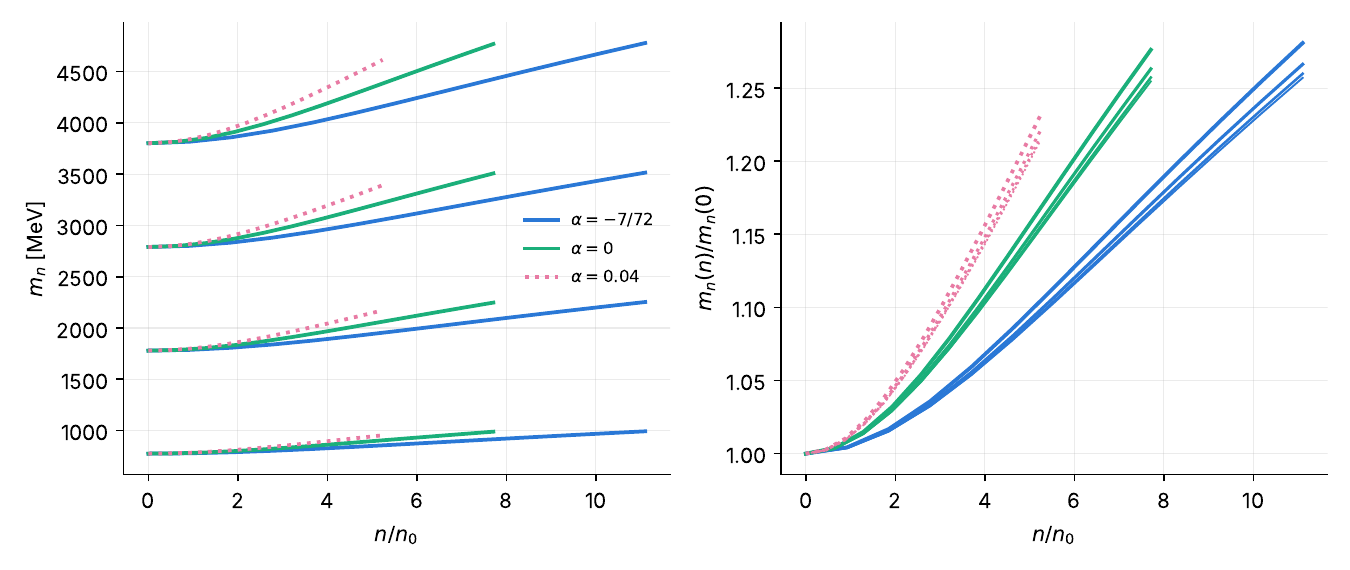}
\caption{Left: masses of the $\rho$ meson and its first three radial excitations in
the confined phase as functions of the quark number density. Right: the same masses
divided by their vacuum values. The four curves for each $\alpha$ nearly coincide.}
\label{fig:masses}
\end{figure}

The masses at finite density are shown in figure~\ref{fig:masses} and
table~\ref{tab:masses}. For $\alpha=0$ the $\rho$ mass increases from $775$~MeV to
$862$~MeV at $n=4n_0$ and to $932$~MeV at $6n_0$. The excited states shift by almost the
same fraction, slightly less than the $\rho$. At fixed density the shift is smaller for
negative $\alpha$. Two effects contribute: the coefficient $c_n$ is smaller, and $\eta$
is smaller because $z_{\rm IR}$ decreases with decreasing $\alpha$. For $\alpha=-7/72$ the $\rho$
mass at $6n_0$ is $874$~MeV. For $\alpha=0.04$ the reality condition
$8\alpha(1+\eta)\le1$ limits the density to $n\lesssim5.3n_0$ at this order.

The quark charge makes $f(z)>f_{\rm AdS}$ near the wall, which shortens the effective size
of the confining region (section~\ref{sec:wkb}) and raises the masses. This differs from
the decrease predicted by Brown--Rho scaling~\cite{Brown:1991kk}. In the present model the
confined phase has no chiral condensate, so the result describes only the effect of the
back-reacted density on the confining geometry.

\begin{table}[tb]
\centering
\begin{tabular}{lccccc}
\toprule
$\alpha$ & $-7/72$ & $-0.05$ & $0$ & $0.02$ & $0.04$\\
\midrule
$c_1$ & 0.281 & 0.296 & 0.321 & 0.336 & 0.355\\
$c_2$ & 0.262 & 0.277 & 0.300 & 0.313 & 0.332\\
$c_3$ & 0.256 & 0.269 & 0.292 & 0.305 & 0.323\\
$c_4$ & 0.253 & 0.267 & 0.289 & 0.302 & 0.320\\
\bottomrule
\end{tabular}
\caption{Coefficients of the small-density expansion
$m_n^2=m_n^2(0)(1+c_n\eta)$ of the vector meson masses.}
\label{tab:cn}
\end{table}

\begin{table}[t]
\centering
\begin{tabular}{lcccc}
\toprule
& $m_1$ & $m_2$ & $m_3$ & $m_4$\\
\midrule
$n=0$ (all $\alpha$) & 775 & 1780 & 2790 & 3801\\
\midrule
$\alpha=-7/72$, $n=2n_0$ & 790 & 1812 & 2839 & 3868\\
$\alpha=-7/72$, $n=4n_0$ & 828 & 1892 & 2962 & 4034\\
$\alpha=-7/72$, $n=6n_0$ & 874 & 1993 & 3116 & 4242\\
\midrule
$\alpha=0$, $n=2n_0$ & 802 & 1837 & 2877 & 3919\\
$\alpha=0$, $n=4n_0$ & 862 & 1968 & 3078 & 4190\\
$\alpha=0$, $n=6n_0$ & 932 & 2120 & 3311 & 4504\\
\midrule
$\alpha=0.04$, $n=2n_0$ & 813 & 1861 & 2915 & 3970\\
$\alpha=0.04$, $n=4n_0$ & 896 & 2043 & 3193 & 4346\\
\bottomrule
\end{tabular}
\caption{Masses in MeV of the $\rho$ meson ($m_1$) and its first three radial
excitations in the dense confined phase.}
\label{tab:masses}
\end{table}

\begin{figure}[t]
\centering
\includegraphics[width=0.66\textwidth]{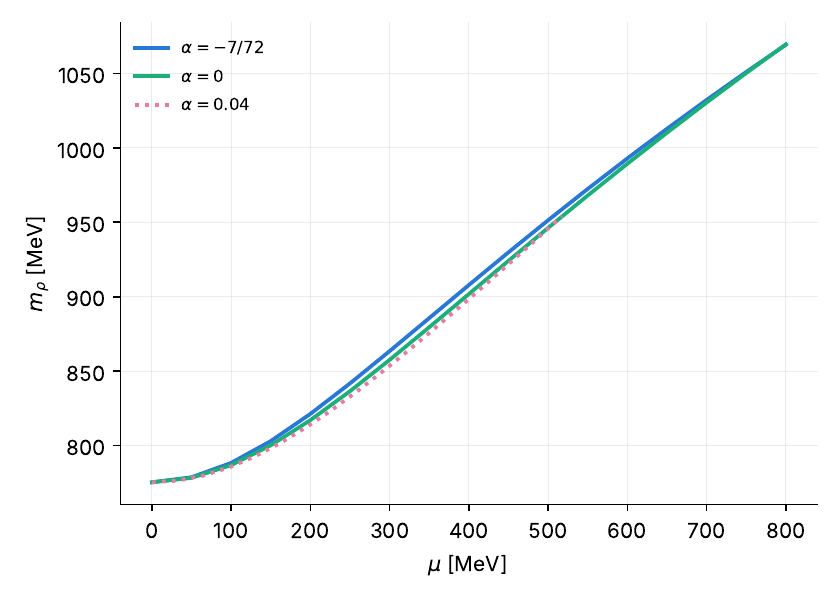}
\caption{Mass of the $\rho$ meson in the confined phase as a function of the quark
chemical potential.}
\label{fig:mrho}
\end{figure}

\begin{figure}[t]
\centering
\includegraphics[width=\textwidth]{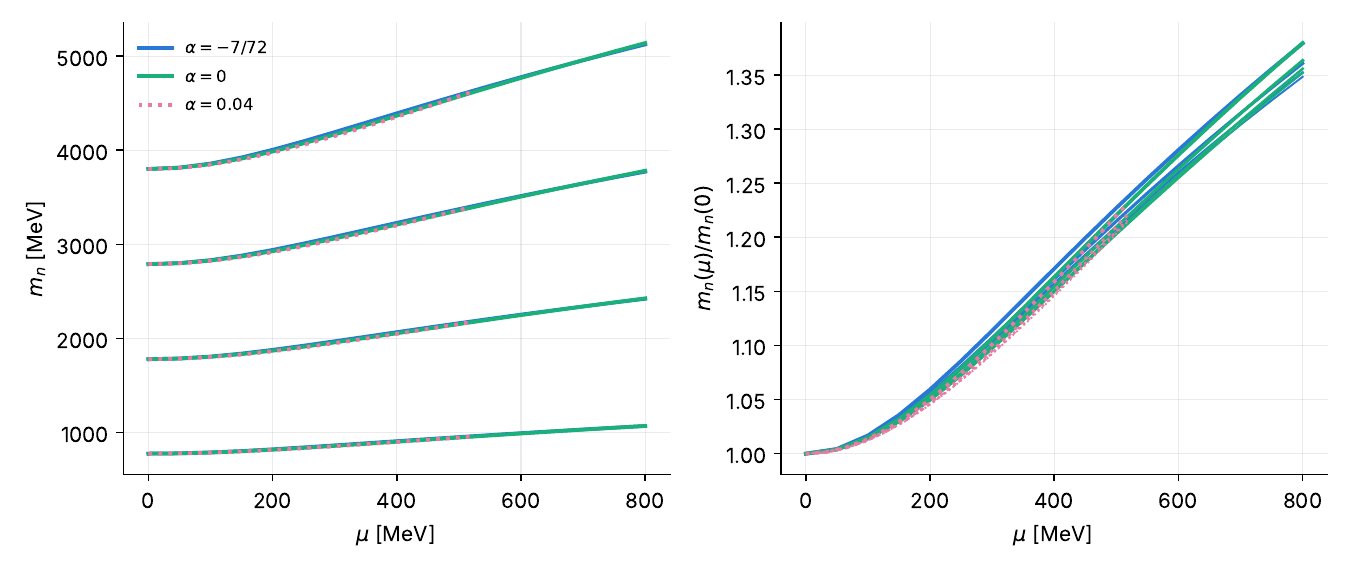}
\caption{Left: masses of the $\rho$ meson and its first three radial excitations as
functions of $\mu$. Right: the same masses divided by their vacuum values.}
\label{fig:mmu}
\end{figure}

\paragraph{Dependence on the chemical potential.}
In the grand canonical ensemble it is natural to express the masses in terms of $\mu$.
With eq.~\eqref{eq:qprime} the density parameter is
\begin{equation}
\eta=\frac{3(1-8\alpha)}{2}\,\frac{N_f}{N_c}\left(\frac{\mu z_{\rm IR}}{A}\right)^2 ,
\end{equation}
so at small $\mu$ the masses grow quadratically,
$m_n\simeq m_n(0)\left[1+\frac{3}{4}c_n (1-8\alpha)\frac{N_f}{N_c}\left(\mu z_{\rm IR}/A\right)^2\right]$.
The $\rho$ mass as a function of $\mu$ is shown in figure~\ref{fig:mrho}, and the first
four masses in figure~\ref{fig:mmu}. At $\alpha=0$, $m_\rho$ is $817$~MeV at
$\mu=200$~MeV, $902$~MeV at $400$~MeV and $1070$~MeV at $800$~MeV. At fixed $\mu$ the
curves for different $\alpha$ nearly coincide: at $\mu=800$~MeV, $m_\rho=1069$~MeV for
$\alpha=-7/72$. Two effects cancel here. For negative $\alpha$ the coefficients $c_n$ and the
cutoff $z_{\rm IR}$ are smaller, which reduces the shift at a given density, but
$1-\zeta=\frac32\sqrt{1-8\alpha}$ is larger, so the confined phase holds more charge at the same $\mu$.
For $\alpha=0.04$ the reality condition stops the curve at $\mu\simeq520$~MeV. The ratio $m_n(\mu)/m_n(0)$ is almost the same for all four modes
(figure~\ref{fig:mmu}, right). These masses refer to the confined phase, which is
stable only below the transition line of figure~\ref{fig:Tcmu}, that is for
$\mu<\mu_0$ at $T=0$.

\subsection{First-order shift for general $\alpha$}

The perturbative result~\eqref{eq:cn} extends to any GB coupling. To first order in
$\eta$ the metric function of tcAdS is $f=f_{\rm AdS}+\eta y^6/\sqrt{1-8\alpha}$. The eigenvalue
$m^2z_{\rm IR}^2$ is the stationary value of the Rayleigh quotient
\begin{equation}
\mathcal R[\psi]=A^2\,\frac{\int_0^1dy\,f\psi'^2/y}{\int_0^1dy\,\psi^2/(fy)} ,
\end{equation}
and varying it with this $\delta f$ gives
\begin{equation}
c_n(\alpha)=\frac{A^2}{\sqrt{1-8\alpha}}\left[
\frac{\int_0^1dy\,y^5\psi_n'^2}{\int_0^1dy\,\psi_n'^2/y}
+\frac{\int_0^1dy\,y^5\psi_n^2}{\int_0^1dy\,\psi_n^2/y}\right],
\qquad \psi_n=y\,J_1(\gamma_{0,n}y).
\label{eq:cnalpha}
\end{equation}
The two ratios are the averages of $y^6$ over the kinetic and potential terms of the
mode. They do not depend on $\alpha$, so the GB coupling enters only through the
prefactor $A^2/\sqrt{1-8\alpha}$. At $\alpha=0$ eq.~\eqref{eq:cnalpha} reduces to eq.~\eqref{eq:cn}. It
reproduces all entries of table~\ref{tab:cn} to four digits.

For highly excited modes the Bessel functions oscillate rapidly, $\psi_n^2$ and
$\psi_n'^2/\gamma_{0,n}^2$ both average to $y/(\pi\gamma_{0,n})$, and each ratio tends to
$\int_0^1dy\,y^6=1/7$. Hence
\begin{equation}
\lim_{n\to\infty}c_n(\alpha)=\frac{2A^2}{7\sqrt{1-8\alpha}},
\label{eq:cinf}
\end{equation}
which equals $2/7=0.286$ at $\alpha=0$, $0.250$ at $\alpha=-7/72$ and $0.316$ at
$\alpha=0.04$. The values in table~\ref{tab:cn} approach this limit from above. The
$\rho$ meson is the most sensitive to density because its wave function is weighted
towards the wall, where $\delta f\propto y^6$ is largest.

\subsection{WKB estimate and the size of the confining region}
\label{sec:wkb}

The density dependence of the whole spectrum has a simple geometric interpretation.
Introduce the coordinate $r$ with $dr=dz/(Af)$. Then eq.~\eqref{eq:vector} becomes
\begin{equation}
\partial_r\!\left(\frac{1}{z}\,\partial_r\psi\right)+\frac{m^2}{z}\,\psi=0 ,
\end{equation}
in which the GB coupling and the density appear only through the map $z(r)$. For large
$m$ the WKB solution with the boundary behaviour of $zJ_1(mz)$ and the Neumann condition
at the wall gives
\begin{equation}
m_n\simeq\frac{\pi\left(n-\frac14\right)}{r_{IR}},\qquad
r_{IR}=\int_0^{z_{\rm IR}}\frac{dz}{A\,f(z)} .
\label{eq:wkb}
\end{equation}
In vacuum $r_{IR}=Az_{\rm IR}$ and eq.~\eqref{eq:wkb} is the large-$n$ form of the zeros
$\gamma_{0,n}\simeq\pi(n-\frac14)$. At finite density $f>f_{\rm AdS}$ near the wall, so $r_{IR}$
is shorter. The masses then rise in the ratio
\begin{equation}
\frac{m_n(\eta)}{m_n(0)}\simeq\left[\int_0^1dy\,\frac{f_{\rm AdS}}{f(y)}\right]^{-1},
\label{eq:wkbratio}
\end{equation}
the same for all highly excited modes. At $\alpha=0$ this is
$\big[\int_0^1dy/(1+\eta y^6)\big]^{-1}\simeq1+\eta/7$, in agreement with
eq.~\eqref{eq:cinf}. Figure~\ref{fig:decay} (right) compares eq.~\eqref{eq:wkbratio}
with the numerical ratio for the fourth mode. The difference is at most $0.3\%$, even at
$\eta\simeq2$.

The quark density therefore shortens the effective size of the confining region. A
direct consequence is that the level spacing grows with density. In vacuum
$m_{n+1}-m_n\simeq\pi/(Az_{\rm IR})=1012$~MeV for all $\alpha$. At $\alpha=0$ the spacing $m_4-m_3$
increases to $1038$~MeV at $1.85\,n_0$, $1118$~MeV at $4.15\,n_0$ and $1193$~MeV at $6\,n_0$.
The spectrum stays linear in $n$, $m_n\propto n$, and does not acquire the Regge form
$m_n^2\propto n$.

\subsection{Decay constants}

\begin{figure}[t]
\centering
\includegraphics[width=\textwidth]{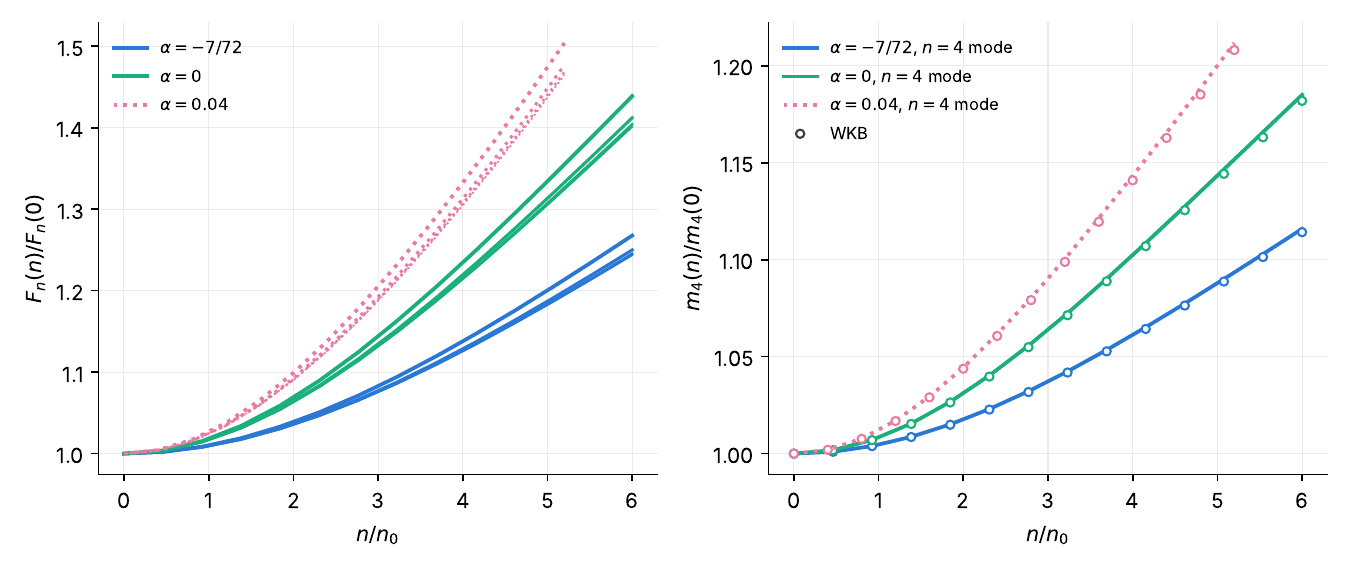}
\caption{Left: decay constants of the $\rho$ meson and its first three radial
excitations in the confined phase, divided by their vacuum values, as functions of
the quark number density. Right: mass ratio of the fourth mode (lines) compared with
the WKB estimate~\eqref{eq:wkbratio} (circles).}
\label{fig:decay}
\end{figure}

The decay constant of the $n$th mode is defined by
$\langle0|J_\mu^a|V_n^b\rangle=F_n\,\delta^{ab}\epsilon_\mu$. In the hard wall model
$F_n=\psi_n''(0)/g_5$~\cite{Erlich:2005qh}, with the wave function normalized by the time
derivative term of the action, $\int_0^{z_{\rm IR}}dz\,\psi_n^2/(A f z)=1$. In vacuum this gives
$F_n\propto\gamma_{0,n}/[J_1(\gamma_{0,n})\,z_{\rm IR}^2]$. The ratio $F_n(n)/F_n(0)$ at fixed $\alpha$
does not depend on the normalization of $g_5$, and we show it in figure~\ref{fig:decay}
(left). At $\alpha=0$ the $\rho$ decay constant grows by $6\%$ at $1.85\,n_0$, $25\%$ at
$4.15\,n_0$ and $44\%$ at $6\,n_0$. The excited states follow closely.

To good accuracy the decay constants scale as the square of the masses,
\begin{equation}
\frac{F_n(n)}{F_n(0)}\simeq\left[\frac{m_n(n)}{m_n(0)}\right]^2 .
\end{equation}
At $\alpha=0$ and $6\,n_0$, for example, the $\rho$ mass ratio is $1.202$ and the decay
constant ratio is $1.439$, compared with $1.202^2=1.444$. This is what one expects if the
density acts mainly through the effective size $r_{IR}$ of the confining region. In
vacuum $F_n$ scales as the inverse square of the wall position, and the masses as its
inverse.

\subsection{Comparison with experiment}

At normal nuclear density the model predicts a small increase of the $\rho$ mass:
$+0.5\%$ for $\alpha=-7/72$, $+0.9\%$ for $\alpha=0$ and $+1.3\%$ for $\alpha=0.04$, that is
$m_\rho=779$, $782$ and $786$~MeV. Experiments on vector mesons in nuclei do not give a
uniform picture; see ref.~\cite{Hayano:2008vn} for a review. The KEK-PS E325
experiment reported a decrease of the $\rho$ and $\omega$ masses by about $9\%$ at normal
nuclear density~\cite{Naruki:2005kd}. The CLAS collaboration found no significant mass
shift of the $\rho$, together with a broadening consistent with collisions in the
nucleus~\cite{Nasseripour:2007aa}. In hot and dense matter produced in In--In
collisions, the NA60 collaboration found a strong broadening of the $\rho$ spectral
function and essentially no mass shift~\cite{Arnaldi:2006jq}.

The shift we obtain at $n_0$ is of order one percent. It is compatible with the
absence of a measurable shift, and it has the opposite sign to the E325 result. At
leading order in $N_c$ the mesons of the model are stable, so the model gives no width
and cannot address the broadening seen by CLAS and NA60. In the confined phase the
masses also do not depend on temperature at this order, because the tcAdS geometry is
the same at all $T$ below the transition. Any temperature dependence of the vector
meson spectrum, and its melting above $T_c$, would require the quasinormal modes of the
black hole, which we leave for future work.

%==================================================================
\section{Conclusion}
\label{sec:conclusion}
%==================================================================

We have studied the confinement\slash deconfinement transition of the hard wall model of
AdS/QCD in Einstein--Maxwell--Gauss--Bonnet gravity at finite quark density. Both phases
include the back-reaction of the quark matter. The deconfined phase is a charged
Gauss--Bonnet black hole and the confined phase is thermally charged AdS with an
infrared cutoff.

The Gauss--Bonnet coupling enters the grand potential of the black hole only through
the normalization of the time direction, and the result satisfies the first law with an
area-law entropy. In the confined phase the value of the gauge field at the infrared
wall is fixed by requiring that the quark number obtained from the boundary term equals
the one obtained from the bulk free energy. This value now depends on the Gauss--Bonnet
coupling and reduces to the one of ref.~\cite{Lee:2009bya} when the coupling vanishes.
The condition holds at second order in the charge, and we used the free energy at that
order.

At fixed quark chemical potential we found the transition temperature in closed form.
At fixed quark number density we generalized the parametric solution of
ref.~\cite{Lee:2009bya} to Gauss--Bonnet gravity. The transition temperature at fixed
density has its maximum where the chemical potentials of the two phases are equal.
Elsewhere on the line of equal free energies the pressures of the two phases differ, so
the physical transition at fixed density passes through a coexistence region, which we
obtained from the transition line at fixed chemical potential. In both ensembles the
transition temperature first rises slightly, because the charged confined phase holds
more quark number than the deconfined phase, and then falls to zero. Its curvature at
small chemical potential therefore has the opposite sign to the lattice value and is
much smaller in magnitude. A negative Gauss--Bonnet coupling raises the transition temperature
and extends the confined phase to higher density. At the lower causality bound the line at fixed chemical potential
ends at 1.65~GeV, compared with 1.23~GeV without the Gauss--Bonnet term, and the
line at fixed density reaches zero temperature at about forty times nuclear density,
compared with about twenty times. A coupling close to $-0.08$ gives a transition temperature of
157~MeV at zero density.

The deconfined phase has the Bekenstein--Hawking entropy and a conformal equation of
state. A negative coupling lowers both the entropy density in units of the temperature
and the quark number susceptibility. The entropy jumps from zero to its black hole value
at the transition, so the transition is of first order in both ensembles. The jump in
quark number density changes sign at the maximum of the transition temperature, in
agreement with the Clausius--Clapeyron relation, which we verified along the numerical
line. Without the Gauss--Bonnet term the fourth root of the latent heat grows from
315~MeV at zero density to about 480~MeV near the end of the line.

In the dense confined phase the masses of the $\rho$ meson and of its radial excitations
increase with density by nearly the same fraction. The rate of increase at small density
falls slowly with the excitation number and grows with the Gauss--Bonnet coupling. In
vacuum the coupling can be absorbed into the infrared cutoff, so the excited states do
not depend on it. They lie above the observed $\rho(1450)$ and $\rho(1700)$, as is usual
for the hard wall. We found the first-order density shift in closed form for any
coupling, together with its limit for highly excited states. A WKB analysis shows that
the density shortens the effective size of the confining region. This raises all masses
by nearly the same factor and increases the level spacing, from 1012~MeV in vacuum to
about 1190~MeV at six times nuclear density. The decay constants grow approximately as
the square of the masses. At normal nuclear density the $\rho$ mass rises by 0.5 to 1.3
percent. This is compatible with the absence of a measurable shift reported by CLAS, and
has the opposite sign to the decrease reported by KEK-PS E325.

Several extensions are possible for above analysis. A chiral sector with a quark condensate would allow the chiral transition to be studied together with deconfinement. A soft wall or dilaton background would give Regge trajectories and could turn the transition into a crossover at small density. The shear viscosity, the electrical conductivity and other transport coefficients can be computed in the same background.

\acknowledgments
S.~Sachan would like to thank Satchi Naik, Sanjay Siwach and Prasanta K. Panigrahi for useful insights and discussions during the project, and Chanyong Park for his help in solving the equations numerically. This work in part is funded by the UP CST project no. CST/D-828.

%==================================================================

\end{document}